\documentclass{aa}  

\usepackage{graphicx}
\usepackage{txfonts}
\usepackage{newtxtext,newtxmath}
\usepackage{graphicx}
\usepackage{xspace}	    
\usepackage{xcolor}
\usepackage{url}
\usepackage{float}
\usepackage{amssymb}

\newcommand{\simgt}{\,\rlap{\lower 3.5 pt \hbox{$\mathchar \sim$}} \raise 1pt \hbox {$>$}\,}
\newcommand{\simlt}{\,\rlap{\lower 3.5 pt \hbox{$\mathchar \sim$}} \raise 1pt \hbox {$<$}\,}

\newcommand{\BE}{\begin{equation}}
\newcommand{\EE}{\end{equation}}
\newcommand{\BEA}{\begin{eqnarray}}
\newcommand{\EEA}{\end{eqnarray}}

\newcommand{\DV}{\ifmmode{\Delta v}\else $\Delta v$\xspace\fi}

\newcommand{\HI}{\ifmmode{\textsc{hi}}\else H\textsc{i}\fi\xspace}
\newcommand{\HII}{\ifmmode{\textsc{hii}}\else H\textsc{ii}\fi\xspace}

\newcommand{\Ha}{\ifmmode{\mathrm{H}\alpha}\else H$\alpha$\fi\xspace}
\newcommand{\OiiiHb}{\ifmmode{\mathrm{[OIII]+H}\beta}\else [OIII]+H$\beta$\fi\xspace}

\newcommand{\Msun}{\ifmmode{M_\odot}\else $M_\odot$\xspace\fi}
\newcommand{\MUV}{\ifmmode{M_\textsc{uv}}\else $M_\textsc{uv}$\xspace\fi}
\newcommand{\fesc}{\ifmmode{f_\mathrm{esc}}\else $f_\mathrm{esc}$\xspace\fi}
\newcommand{\lya}{\ifmmode{\mathrm{Ly}\alpha}\else Ly$\alpha$\xspace\fi}

\newcommand{\nh}[1][]{\ifmmode{\overline{n}_\textsc{h}^{#1}}\else $\overline{n}_\textsc{h}$\xspace\fi}

\newcommand{\Wobs}{\ifmmode{EW_\textrm{obs}}\else $EW_\textrm{obs}$\xspace\fi}
\newcommand{\Wint}{\ifmmode{EW_\textrm{int}}\else $EW_\textrm{int}$\xspace\fi}

\newcommand{\xHI}{\ifmmode{\overline{x}_\HI}\else $\overline{x}_\HI$\xspace\fi}
\newcommand{\trec}{\ifmmode{t_\textrm{rec}}\else $t_\textrm{rec}$\xspace\fi}
\newcommand{\clump}[1][]{\ifmmode{C_\HII^{#1}}\else $C_\HII$\xspace\fi}
\newcommand{\xiion}{\ifmmode{\xi_\mathrm{ion}}\else $\xi_\mathrm{ion}$\xspace\fi}

\newcommand{\Nion}{\ifmmode{\dot{N}_{\mathrm{ion}}}\else $\dot{N}_\mathrm{ion}$\xspace\fi}
\newcommand{\Rion}[1][]{\ifmmode{R_\mathrm{ion}^{#1}} \else $R_\mathrm{ion}$\xspace\fi}

\newcommand{\Rchar}{\ifmmode{R_{\mathrm{char}}}\else $R_\mathrm{char}$\xspace\fi}
\newcommand{\Rbub}{\ifmmode{R_{\mathrm{ion}}}\else $R_\mathrm{ion}$\xspace\fi}
\newcommand{\Rexp}{\ifmmode{\langle R \rangle}\else $\langle R \rangle$\xspace\fi}
\newcommand{\dpdlogR}{\ifmmode{dp/d\log_{10}R}\else $dp/d\log_{10}R$\xspace\fi}
\newcommand{\od}{\ifmmode{N/\langle N \rangle}\else $N/\langle N \rangle$\xspace\fi}

\newcommand{\kms}{\,\ifmmode{\mathrm{km}\,\mathrm{s}^{-1}}\else km\,s${}^{-1}$\fi\xspace}
\newcommand{\cm}{\,\ifmmode{\mathrm{cm}}\else cm\fi\xspace}

\usepackage{hyperref}
\hypersetup{
    colorlinks=true,
    linkcolor=red,
    filecolor=black, 
    citecolor=blue,
    urlcolor=magenta,
    }
\begin{document}

   \title{Mapping reionization bubbles in the JWST era I: \\
   empirical edge detection with Lyman alpha emission from galaxies}
   \titlerunning{Mapping reionization bubbles in the JWST era I}


   \author{Ting-Yi Lu \inst{1, 2}
          \and
          Charlotte A. Mason \inst{1, 2}
          \and
          Andrei Mesinger \inst{3}
          \and
          David Prelogović \inst{3}
          \and
          Ivan Nikolić \inst{3}
          \and 
          Anne Hutter \inst{1, 2}
          \and \\
          Samuel Gagnon-Hartman  \inst{3}
          \and 
          Mengtao Tang  \inst{4}
          \and 
          Yuxiang Qin \inst{5}
          \and          
          Koki Kakiichi \inst{1, 2}
          }

   \institute{Cosmic Dawn Center (DAWN)
         \and
             Niels Bohr Institute, University of Copenhagen, Jagtvej 128, 2200 Copenhagen N, Denmark
         \and
             Scuola Normale Superiore, Piazza dei Cavalieri 7, I-56126 Pisa, Italy
         \and
             Steward Observatory, University of Arizona, 933 N Cherry Ave, Tucson, AZ 85721, USA
         \and
             Research School of Astronomy and Astrophysics, Australian National University, Canberra, ACT 2611, Australia
             }

   \date{Received; accepted}

 
  \abstract 
   {Ionized bubble sizes during reionization trace physical properties of the first galaxies. 
   JWST's ability to spectroscopically confirm and measure Lyman-alpha (\lya) emission in sub-L$^\star$ galaxies opens the door to mapping ionized bubbles in 3D. However, existing \lya-based bubble measurement strategies rely on constraints from single galaxies, which are limited by the large variability in intrinsic \lya emission.}
   {As a first step, we present two bubble size estimation methods using \lya spectroscopy of \textit{ensembles} of galaxies, enabling us to map ionized structures and marginalize over \lya emission variability. We test our methods using Gpc-scale reionization simulations of the intergalactic medium (IGM).}
   {To map bubbles in the plane of the sky, we develop an edge detection method based on the asymmetry of \lya transmission as a function of spatial position. To map bubbles along the line-of-sight, we develop an algorithm using the tight relation between \lya transmission and the line-of-sight distance from galaxies to the nearest neutral IGM patch.}
   {Both methods can robustly recover bubbles with radius $\simgt$10 comoving Mpc, sufficient for mapping bubbles even in the early phases of reionization, when the IGM is $\sim70-90\%$ neutral. These methods require $\simgt$0.002-0.004 galaxies/cMpc$^3$, a 5$\sigma$ \lya equivalent width upper limit of $\lesssim$30\,\AA\, for the faintest targets, and redshift precision $\Delta z \simlt 0.015$, feasible with JWST spectroscopy. Shallower observations will provide robust lower limits on bubble sizes. Additional constraints on IGM transmission from \lya escape fractions and line profiles will further refine these methods, paving the way to our first direct understanding of ionized bubble growth.}
  {}


   \maketitle
%

\section{Introduction}  \label{sec:intro}

    Between $20\gtrsim z \gtrsim 5$, the first luminous sources reionized neutral hydrogen in the intergalactic medium (IGM), forming ionized bubbles around them which eventually merged \citep[e.g., see][for a recent review]{Mesinger2016}. 
    Understanding the properties, formation, and evolution of these first sources and how they drove the reionization process is a frontier in astronomy. 
    While significant progress has been made in the last two decades in measuring the volumn-averaged timeline of reionization \citep[e.g.,][]{Fan2006,Stark2010,Treu2012,Pentericci2014,Mason2018,Davies2018b,jung_texas_2020,Qin2021b,Bosman2022,Bolan2022,Umeda2023,Nakane2023,Tang2024c}, we currently have no robust measurements of the typical sizes of ionized bubbles and how they evolve over time. 

    The first large bubbles are predicted to form around early galaxy overdensities, thus the sizes of ionized regions and their host overdensities can provide key insights into the reionization process. In individual regions this would enable ionizing photon `accounting', where visible galaxies' ionizing photon output can be compared to that required to grow the bubble, providing insight into their ionizing photon escape fractions and star formation histories, and the contribution of galaxies below even JWST's detection limits \citep[e.g.,][]{Shapiro1987,Haiman2002,Mason2020,endsley_strong_2022, Whitler2024,Torralba-Torregrosa2024}.
    On a global scale, the distribution of ionized region sizes at fixed redshift -- the reionization `morphology' -- is sensitive to the halo mass scale of the dominant ionizing sources \citep[e.g.,][]{McQuinn2007a, Mesinger2016, Seiler2019a, Hutter2023, Lu2024}, with models where low mass halos ($M_h \sim 10^8\,\Msun$) dominate the ionizing photon budget predicting median bubble radii of $\sim10-30$\,cMpc ($\sim4-12$ arcmin) during the mid-stages of reionization. 
    In the next decade, there will be substantial observational efforts to identify and start characterizing the sizes of ionized regions over scales large enough to capture multiple bubbles, via upcoming 21cm experiments to observe the neutral hydrogen directly \citep[e.g.][]{Yatawatta2013, Bowman2013} and wide-area near-IR spectroscopic surveys to observe galaxies which should trace ionized regions e.g. with Euclid, Roman and Subaru/PFS \citep{Euclid_deepz6_2022,Euclid_wide2022,Wang2022,Greene2022_PFS}.
    
    Lyman-alpha ($\lya$) absorption by neutral hydrogen in high-$z$ objects such as galaxies, QSOs, and gamma-ray bursts, is currently our best tool to identify ionized regions, as \lya suffers strong damping wing absorption due to neutral hydrogen in the foreground of the objects \citep{Miralda-Escude1998a, Mesinger2008}. Although considerably fainter than QSOs, galaxies have the benefit of being more numerous and less biased tracers of the EoR morphology \citep{Shen2007,Eilers2024},
    such that they can be used to sample the IGM. 
    Spectroscopic and narrow-band surveys in the last decades have identified a number of fields showing an excess of \lya emission from galaxies at $z\sim6-9$, implying large-scale ($R>1$\,pMpc) ionized regions \citep[e.g.,][]{Castellano2016b,Tilvi2020,endsley_strong_2022,Wold2022,Napolitano2024,Tang2024c}. Upcoming wide-area near-IR spectroscopy with Euclid, Roman and PFS will be able to identify more candidate ionized regions by detecting bright ($\simgt 1\times10^{-17}$\,erg s$^{-1}$ cm$^{-2}$) \lya emission, and photometric overdensities of UV-bright galaxies. However, as we will demonstrate, these instruments will not have the sensitivity to sample, nor wavelength coverage to spectroscopically confirm, sufficient numbers of galaxies in these ionized regions to fully map bubbles and characterize their sizes \citep[][]{Hutter2018,Zackrisson2020}, thus deeper follow-up imaging and spectroscopy will be required.

  JWST has provided the unique capabilities to make \textit{mapping} these ionized regions, and the overdensities they surround, feasible. Its ability to spectroscopically confirm UV faint $z>6$ galaxies, even down to \MUV$\approx-18$ at $z\sim14$ \citep[][]{Curtis-Lake2023,Carniani2024} is allowing us to precisely map $z>5$ galaxy distributions in 3D for the first time \citep[e.g.,][]{Helton2024,Meyer2024,Chen2024,Tang2024c}. The sensitivity of NIRSpec \citep{Jakobsen2022} has enabled the detection \lya emission from UV-faint galaxies \citep[e.g.,][]{Saxena2023,Tang2023,Chen2024}, which has been infeasible from the ground, and can be used to comprehensively trace ionized regions. JWST's ability to measure \lya escape fractions from Balmer lines also provides a promising new indicator of highly ionized regions, where the fraction of \lya escaping from galaxies through the IGM can reach close to unity \citep[e.g.,][]{Chen2024}. Such power opens a new window for mapping ionized bubbles. 
This potential has been demonstrated empirically in early JWST results. For example, in the CEERS/EGS field -- which contains $\simgt50\%$ of all \lya emission known to-date above $z>7$ and is likely the largest known ionized region at this epoch \citep{Oesch2015,Roberts-Borsani2016,Stark2017,Tilvi2020,Jung2022,Larson2022}. Several studies using NIRSpec spectroscopy have detected new high equivalent width (EW$\simgt100$\,\AA) \lya-emission, and measured \lya escape fractions to confirm several galaxies in the region likely transmit $>50\%$ of their \lya through the IGM \citep{Jung2023,Tang2023,Tang2024c,Napolitano2024,Chen2024}.
\citet{Chen2024} and \citet{Tang2024c} demonstrated $z\sim7-8$ galaxies in EGS appear clustered in several narrow redshift bins around \lya emitting galaxies, hinting at several pMpc-scale bubbles along the line of sight.

  However, a quantitative link from discrete galaxy observations such as these to the sizes of ionized regions has so far been challenging due to the intrinsic variability of \lya\ emission. With an estimate of the fraction of \lya transmitted through the IGM, it should be possible to infer the line-of-sight distance of a galaxy to neutral gas \citep[e.g.,][]{Mesinger2008,Mason2020}, providing we can estimate the \lya\ emitted by the source, \textit{before} attenuation by the IGM. 
  In the past few years, significant progress has been made in understanding the properties of \lya emerging from post-reionization ($5\lesssim z\lesssim6$) galaxies \citep[e.g.][]{Prieto-Lyon2023b, Tang2024b, Bolan2024,Lin2024}. In particular, JWST observations provide key information about the relations between galaxies' \lya emission (rest-frame equivalent widths, \lya escape fractions $f_{\rm esc, Ly\alpha}$, and \lya velocity offsets and line-profiles) and galaxy UV and optical properties such as UV continuum magnitude, UV slopes and [OIII]+H$\beta$ emission at the end of reionization, $z\sim5-6$ \citep[e.g.][]{Chen2024, Tang2024b,Saxena2023, Lin2024,Prieto-Lyon2023b}. These observations can be used as baselines to infer the impact of the neutral IGM on \lya during reionization. 
  However, even in the ionized IGM, \lya\ properties such as EW, escape fraction and lineshape can vary by orders of magnitude in galaxies with similar physical properties \citep[e.g.,][]{Stark2010,Oyarzun2017,Jaskot2019,Hayes2023a}, which radiative transfer simulations predict is largely due to sightline variability in neutral gas and dust distributions in the ISM and CGM \citep[e.g.,][]{Smith2017,Blaizot2023}.

The first attempts to infer the sizes of ionized regions have focused on individual galaxies, using the approach described above to infer the relation between \lya transmission and the distance of the galaxy to the neutral IGM \citep[e.g.,][]{Mason2020,Hayes2023b,Umeda2023,Witstok2024,Torralba-Torregrosa2024}. However, the intrinsic variability of \lya emission makes estimates of the sizes of ionized bubbles based on single galaxies highly uncertain. To provide robust estimates of the sizes of ionized regions thus requires combining information from \textit{ensembles} of galaxies to overcome this variability.

  In this work, we thus aim at providing simple approaches to map ionized bubbles using observations of ensembles of galaxies, given the new capabilities of JWST. Here we focus on empirical methods for detecting bubble edges via integrated \lya observations (i.e. EW, escape fractions). In a companion paper, Nikolić et al in prep, we will present a framework for inferring the centers and radii of individual ionized regions based on \lya line profiles. In Section~\ref{sec:method} we introduce our IGM simulations and forward modeling of \lya emission. We develop methods to map ionized bubbles on the plane of the sky (Section~\ref{sec:asymscore}) and along sightline skewers (Section~\ref{sec:LOSbubble}) using galaxy \lya emission observations, and describe the observational requirements for robust bubble size recovery. Our methods make use of the almost one-to-one relation between the IGM damping wing absorption profile and the distance of galaxies to the nearest neutral IGM. In Section~\ref{sec:discussion}, we discuss prospect for ionized bubble mapping with current and future facilities and surveys. We present our conclusions in Section~\ref{sec:conclusions}.
  
  Throughout this paper, we use a flat $\Lambda$CDM cosmology with $\Omega_{\rm m}=0.31$, $\Omega_{\rm \Lambda}=0.69$, and $h=0.68$ and magnitudes are in the AB system.

\section{IGM simulations and forward modeling galaxy observables} \label{sec:method}

We create simulations of the IGM with forward-modeled galaxy observables to test our bubble mapping methods. In this section, we describe how we simulate ionization cubes (Section~\ref{sec:21cmfast}), model emitted \lya lines from galaxies (Section~\ref{sec:galmod}) and infer the fraction of \lya flux transmitted through the IGM for each galaxy (Section~\ref{methods:T}), which is the key input to our bubble mapping methods.

\subsection{IGM simulations}\label{sec:21cmfast}
We use \texttt{21cmFASTv2}\footnote{\url{https://github.com/andreimesinger/21cmFAST}} \citep{Mesinger2011, Sobacchi2014a, Mesinger2016} to create realistic, large volume reionization IGM simulations. \texttt{21cmFASTv2} is a semi-numerical code that can quickly create reionization simulations for coeval density cubes \citep{Mesinger2011}. 

We create a (300 cMpc)$^{3}$ density cube with (1.5 cMpc)$^{3}$ resolution. We generate ionization cubes of different \xHI by changing the ionization efficiency of galaxies, similar to the approach described by \citet{Lu2024}. For simplicity, we use a bright-galaxy-driven reionization model, in which $M_{h}>10^{11}M_{_\odot}$ halos can ionize the IGM. The broad bubble size distribution of our bright-galaxy-driven reionization model allows us to test bubble size recovery for a wide range of bubble size. As we find our results do not depend significantly on \xHI (which is the dominant driver of bubble sizes) we do not expect our results to change significantly if we had assumed another reionization model.

We extract a catalog of halo masses and positions from the density cube using extended Press-Schechter theory \citep{Sheth2001} and a halo-filtering method developed by \citet{Mesinger2007}. The minimum halo mass extracted in this work is $\sim 4.2 \times 10^{9}M_{\odot}$. As described in \citet{Lu2024}, we use the UV luminosity - halo mass relation from \citet{Mason2015,Mason2022}, adding a scatter of 0.5\,mag following \citet{Ren2018,Whitler2020}, to map from halo masses to UV magnitude. \citet{Lu2024} demonstrated this reproduces the observed UV LFs very well at $z\sim7-10$, though at the faint end (\MUV$\gtrsim$-18) of LF in our simulation slightly underpredicts the number density of galaxies. In later sections we will investigate bubble mapping given different galaxy number densities by setting different \MUV detection limits for mock observations with our simulations. When discussing the corresponding detection limits in real observations, we will quote \MUV limits which match our simulated number densities using the LFs in \cite{Bouwens2021}.

\subsection{Forward modelling \lya emission}\label{sec:galmod}
We will use galaxies' \lya emission properties to recover transmissions at their positions in a bubble. To obtain these properties, we forward model the observed Lya emission lines following the approach by \citet{Mason2018}. We assume the intrinsic \lya\ line shape is a single Gaussian, and model the observed line after transfer through interstellar medium (ISM), circumgalactic medium (CGM), and IGM.

We generate the \lya damping wing optical depth due to the IGM for every halo (selecting $\sim10^5$ halos with $\MUV\leq-17$) along a sightline in the $+z$-direction from the galaxy to a distance of 200 cMpc away\footnote{Beyond this distance, HI contributes negligibly to damping wing absorption. \citep{Mesinger2008}}, using periodic boundary conditions. We calculate the \lya damping wing optical depth profile, $\tau_{\rm IGM}(\Delta v_{\rm \lya})$ \cite[e.g.,][]{Miralda-Escude1998b} at \lya velocity offsets from systemic ($\Delta v_{\rm \lya}$) between -300 km/s and 5700 km/s by summing the contributions of line-of-sight (LOS) neutral hydrogen in the ionization cube as described in \cite{Mesinger2015}.

Our Gaussian \lya emission line model is described by two quantities: the velocity offset from systemic ($\Delta v_{\rm Ly\alpha}$) and line width ($\sigma_{\rm v}$).
Following \citet{Mason2018}, we assume the velocity offset of \lya emission depends on a galaxy's mass. Velocity offsets are observed to be positively correlated with quantities which scale with galaxy mass, likely because of scattering in the ISM and outflows \citep[e.g.][]{Yang2017, Garel2021, Blaizot2023, Hayes2023a}. We sample $\Delta v_{\rm Ly\alpha}$ from the $P(\Delta v_{\rm Ly\alpha}|M_{h})$ distributions in \citet{Mason2018} for each galaxy.
The line width, $\sigma_{\rm v}$, also depends on scattering in ISM and is thus related to $\Delta v_{\rm Ly\alpha}$ \citep{Neufeld1991}. Motivated by the close relationship between $\Delta v_{\rm Ly\alpha}$ and \lya FWHM observed by \citet{Verhamme2015} we model it using $\sigma_{\rm v}=20\mathrm{\,km/s}+\Delta v_{\rm Ly\alpha}/2.355$ (adding a small constant to avoid zero dispersion).

To account for absorption due to infalling residual neutral gas in the ionized IGM/CGM, which is expected to suppress \lya\ flux redward of systemic at $z>5$  \citep{Santos2004a, Dijkstra2011, Mason2018,Park2021}, we cut off \lya emission blueward of the circular velocity ($v_{\rm circ}= [10GM_{\rm h}H(z)]^{1/3}$). We define the emitted \lya line profile, $J_{\alpha}(\Delta v_{\rm Ly\alpha}, M_{h}, v)$, as the \lya line after CGM absorption and assume ISM and CGM properties do not evolve significantly with redshift at $z\sim6-9$. As our methods rely on the relative transmission between sources at the same redshift, redshift evolution of \lya emission or CGM properties should not significantly impact our results.

We define the integrated \lya transmission, $\mathcal{T}$, for each galaxy as the ratio between the observed and emitted line equivalent width (EW):
\begin{equation}\label{eq:T_EWobs_EWem}
    \mathcal{T}=\frac{EW_{\rm \lya,  obs}}{EW_{\rm \lya,  em}} \equiv \frac{\int^{\infty}_{-\infty}dv J_{\alpha}(\Delta v_{\rm \lya}, M_{h}, v)e^{-\tau_{\rm IGM}(\Delta v_{\rm \lya})}}{\int^{\infty}_{-\infty}dv J_{\alpha}(\Delta v_{\rm \lya}, M_{h}, v)}.
\end{equation}

In the following sections, we will refer to this integrated \lya transmission as "\lya transmission".

To model the emitted \lya EW ($EW_{\rm \lya,  em}$) we use an empirically determined relation between galaxy \MUV and observed \lya equivalent width (\Wobs) at $z\sim5-6$, derived by \citet{Tang2024c}. These distributions are based on $>700$ $z\sim5-6$ Lyman-break galaxies with ground-based \lya spectroscopy and JWST photometry, and should be representative of the \lya emission before significant IGM damping wing attenuation, including \lya transmission in the ISM and CGM, i.e. including the impact of local damped \lya absorbers \citep[e.g.,][]{Shapley2003,Reddy2016,Heintz2024}.
We draw the emitted \lya equivalent width, $EW_{ em}$, from the P($EW_{ em}|$\MUV) distributions derived by \citet{Tang2024b}, using the \MUV for each halo as described in Section~\ref{sec:21cmfast}.

\subsection{Estimating \lya transmission from (mock) observations} \label{methods:T}

Our forward model of galaxy \lya\ spectra enables us to create mock observations. From these we now seek to \textit{recover} the \lya transmission, $\mathcal{T}$, for each galaxy to use as a probe of the IGM.
We can use $EW_\mathrm{obs}$ of a galaxy to recover $\mathcal{T}$ given Equation~\ref{eq:T_EWobs_EWem}. Using Bayes' theorem, the inferred probability distribution of $\mathcal{T}$ for a galaxy with an observed \MUV and $EW_\mathrm{obs}$ can be recovered as
\begin{equation}
    P(\mathcal{T}|\MUV, EW_\mathrm{obs}) \propto P(EW_\mathrm{obs}|\mathcal{T},\MUV)P(\mathcal{T}), 
\end{equation}
where $P(EW_\mathrm{obs}|\mathcal{T},\MUV)$ is the likelihood of obtaining $EW_\mathrm{obs}$ given a transmission fraction $\mathcal{T}$ and galaxy's UV magnitude \MUV. $P(\mathcal{T})$ is the prior on \lya transmission. Technically, $\mathcal{T}$ can depend on $\MUV$ because of environmental effects, e.g. brighter galaxies may be more likely to reside in overdense, and thus more ionized, regions \citep{Mason2018b, Mason2018}. However, as we are interested in galaxies which are co-located spatially, we can assume they have similar environments, and thus we do not include any dependence of $\mathcal{T}$ on \MUV. We will explore a more complete treatment in future work.
We can build the likelihood $P(EW_\mathrm{obs}|\mathcal{T},\MUV)$ using $P_0(EW)$, our `intrinsic' EW distribution at $z\sim5-6$ (see Section~\ref{sec:galmod} and \citet{Tang2024b}):
$P(EW_\mathrm{obs}|\mathcal{T},\MUV)\propto P_0(EW_\mathrm{obs}/\mathcal{T}\,|\,\MUV)$.

To account uncertainties or upper limits in $EW_\mathrm{obs}$ measurements, we marginalize over $EW_\mathrm{obs}$ \citep[e.g.,][]{Schenker2014,Mason2018,Tang2024c}. For a galaxy with detected $EW_\mathrm{obs,noise}$ and uncertainty, $\sigma_{EW}$, we assume a Gaussian $EW_\mathrm{obs}$ distribution, such that the inferred $\mathcal{T}$ distribution can be written
\begin{multline}
        P(\mathcal{T}|\MUV, EW_\mathrm{obs,noise}) \propto P(\mathcal{T})\times \\
\int_0^\infty \frac{\exp\left[{-\frac{\left(EW_\mathrm{obs}-EW_\mathrm{obs,noise}\right)^2}{2\sigma_{EW}^2}}\right]}{\sqrt{2\pi}\sigma_{EW}}P(EW_\mathrm{obs}|\mathcal{T},\MUV)dEW_\mathrm{obs}.
\end{multline}

For a galaxy with \lya equivalent width below the $5\sigma$ detection limit, the inferred $\mathcal{T}$ distribution can be written 
\begin{multline}
        P(\mathcal{T}|\MUV, EW_\mathrm{obs,noise} < 5\sigma_{EW}) \propto P(\mathcal{T})\times \\
        \frac{1}{2}\int_{0}^{\infty}\mathrm{erfc}\left(\frac{EW_\mathrm{obs}-5\sigma_{EW}}{\sqrt{2}\sigma_{EW}}\right)P(EW_\mathrm{obs}|\mathcal{T},\MUV)dEW_\mathrm{obs}.
\end{multline}

We assume a uniform prior on $p(\mathcal{T})$ between 0 and 1. We note that \lya\ escape fractions estimated from Balmer line observations could be used as a prior on $p(\mathcal{T})$, which would further improve our estimates of the IGM transmission (e.g. Nikolić et al. in prep), but for simplicity here we consider only measurements of the \lya EW.

Due to the large dispersion of the emitted EW distribution \citep{Tang2024b}, the inferred $P(\mathcal{T}|\MUV, EW_\mathrm{obs,noise})$ for individual galaxies can be very broad.
By making use of ensembles of galaxies, our method marginalizes over this intrinsic variability to recover bubble sizes by locating positions where many galaxies' observed \lya properties are consistent with high $\mathcal{T}$. 

As our method relies on measuring changes in transmission relative to that expected in an ionized IGM this sets an approximate maximum required \lya EW sensitivity limit comparable to the median EW for galaxies in a fully ionized IGM \citep[$\approx30$\,\AA, e.g.,][]{Stark2011,jung_texas_2020,Tang2023}, i.e. such that we can test if the observed EW distribution is consistent with $\mathcal{T}\approx1$ (inside a large bubble) or if $\mathcal{T}$ is significantly lower (in small bubbles or the neutral IGM).

\section{Mapping bubbles in the plane of the sky}\label{sec:asymscore}
\begin{figure*}
    \centering
    \includegraphics[width=0.9\textwidth]{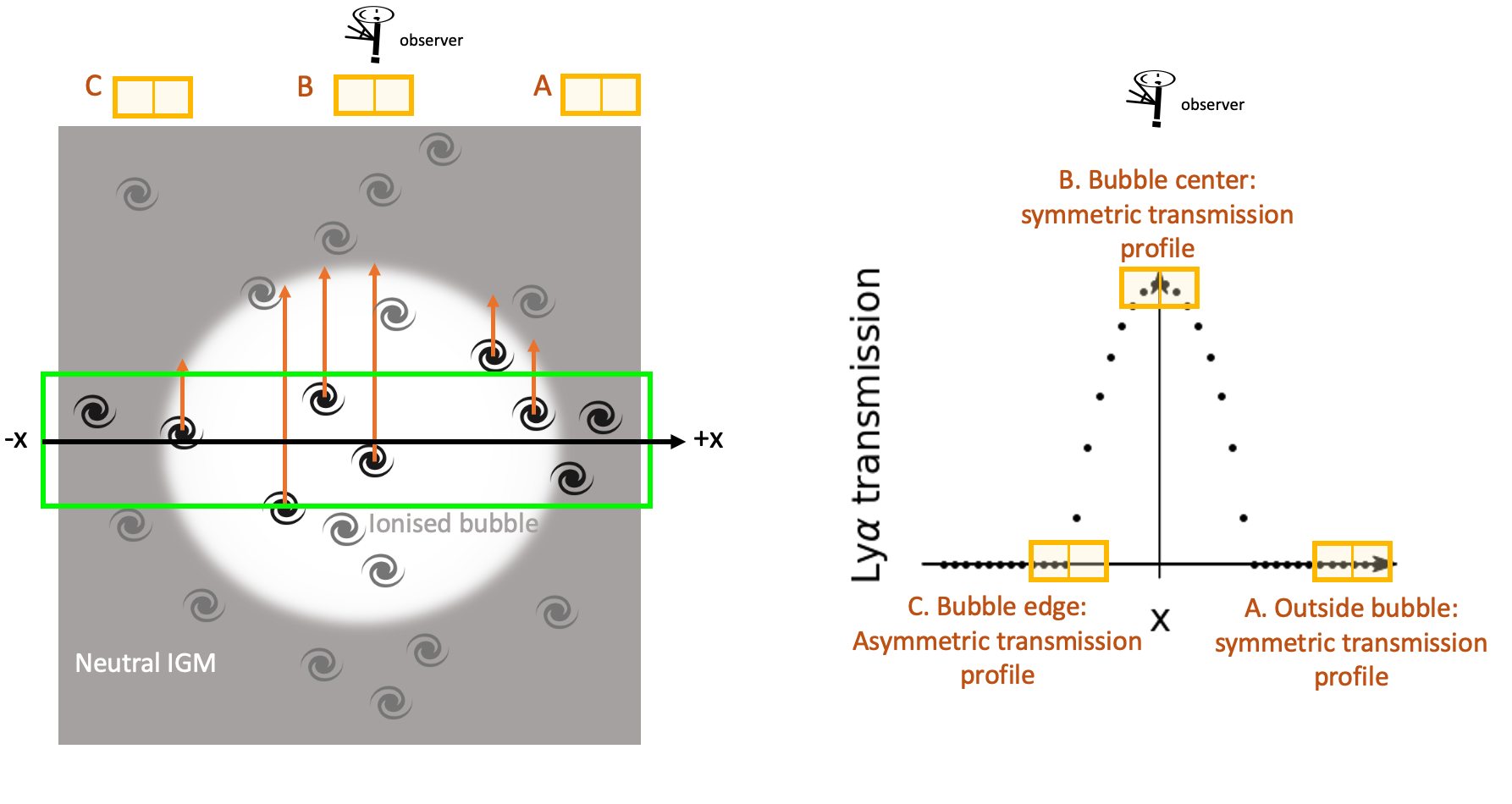}
    \caption{A cartoon to illustrate the \lya transmission profile of galaxies in a plane of the sky around an ionized bubble. \textit{Left}: galaxies in and around a spherical bubble with the observer at the top. \textit{Right}: The integrated \lya transmissions, $\mathcal{T}$ as a function of transverse distance ($x$) measured using galaxies in the green box in the left panel. Transmission profiles $\mathcal{T}(x)$ viewed in different windows (A, B, and C) are different. \textbf{Window A:} Galaxies are all outside the bubble and all have $\mathcal{T}\approx0$, resulting in a flat transmission profile in the window. \textbf{Window B}: Galaxies are located around the center of the bubble, therefore $\mathcal{T}$ in the window are all high and the transmission profile is symmetric around the window center. \textbf{Window B}: Half of the galaxies are outside the bubble and have $\mathcal{T}\approx0$. The other half of the galaxies are inside the bubble, having rising transmission towards the bubble center. The transmission profile around the center of window C is therefore asymmetric. The level of asymmetry can therefore be a probe of bubble edges.  }
    \label{fig:toyasym}
\end{figure*}

\begin{figure*}
    \centering
    \includegraphics[width=\textwidth]{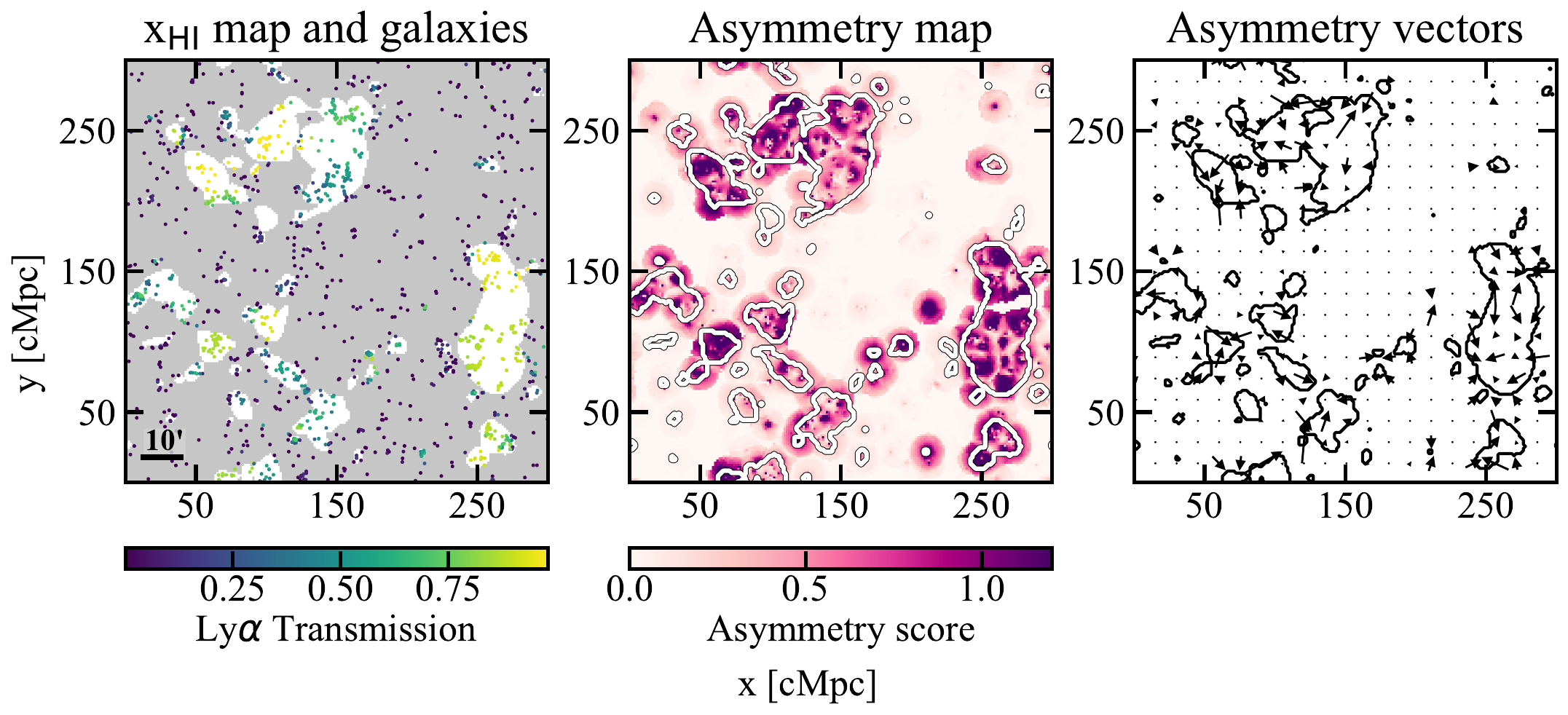}
    \caption{An example of an asymmetry score map made using galaxies' \lya transmission. \textit{Left:} Simulated ionization map (4 cMpc slice from our simulation with mean neutral fraction $\xHI= 0.8$) and galaxies used to calculate asymmetry score. White regions are ionized, grey are neutral. The galaxies are color-coded by their modeled \lya transmission, where we see galaxies with highest transmission are in the centers of bubbles. In the lower left corner we display a 10' ruler (at $z=8$) for gauging bubble size in angular scale. \textit{Middle:} Asymmetry score map (red map). We plot the input ionization map (white contours) on top of the map. As expected, bubble edges have higher asymmetry score. \textit{Right:} Asymmetry vector map where we see asymmetry vectors, color-coded by asymmetry score, point toward bubble edges, black contours mark the input ionization map. }
    \label{fig:exampasymmap}
\end{figure*}

\begin{figure*}
    \centering
    \includegraphics[width=\textwidth]{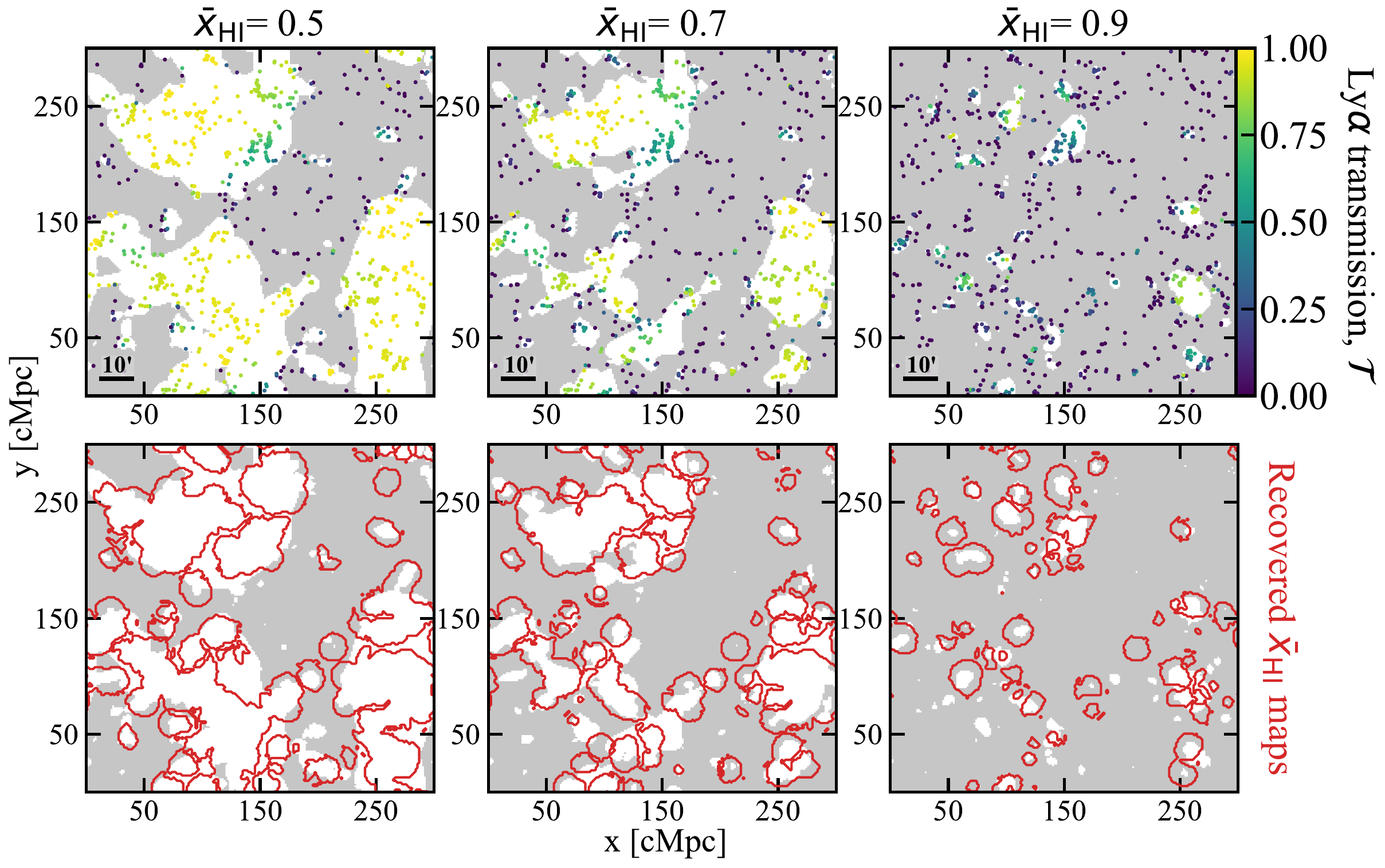}
    \caption{Top panels: Ionization maps and galaxies colored by their \lya transmissions, at \xHI=0.5, 0.7, and 0.9 for our $z=8$ simulation. Lower panels: Recovered bubble edges (red contours) using the approach described in Section~\ref{sec:asymscore} and Appendix~\ref{sec:binarymap} on top of the input ionization maps (white -- ionized, grey -- neutral). For our mock observations we use a number density of $n_\mathrm{gal}=0.002/$cMpc$^3$ (corresponding to a detection limit of \MUV$<-17.8$ at $z=8$) and a \lya\ 5$\sigma$ equivalent width limit of $\geq$30\,\AA\ for the faintest galaxies. We see the bubble edges are recovered well at all neutral fractions.}
    \label{fig:asymmap_2cases_3xHI}
\end{figure*}

Here we describe our method of making bubble maps in the plane of the sky, using an edge-detection technique based on the asymmetric distribution of \lya transmission around the edges of ionized bubbles (Section~\ref{sec:asymscoremethod}). We detail the suggested parameter setup for our method in Section~\ref{sec:bestwindowrnd_xHI} and the observational requirements in Section~\ref{sec:asymmap_require}. 

\subsection{Transmission asymmetry score}\label{sec:asymscoremethod}
In the plane of the sky, we expect galaxies inside an ionized bubble to have higher \lya transmission compared to galaxies outside the bubble -- i.e. bubbles should be signposted by strong \lya-emitters. To map the bubbles requires identifying the edges, which should be marked by a steep drop in \lya transmission. At a point on the sky we can calculate how `asymmetric' \lya transmission is as a function of transverse distance, and identify bubble edges as places where this asymmetry is maximized. In principle, we can also define a threshold in transmission to define bubble edges \citep[as suggested by][]{Kakiichi2023}, but this transmission threshold will depend on the size of the bubbles and thus the average IGM neutral fraction, \xHI. Thus here we consider the \textit{asymmetry} of transmission as it should be more independent of bubble sizes and the reionization history (as we discuss below), making it a more flexible technique.

In Figure~\ref{fig:toyasym} we show a toy model for our method assuming a spherical ionized bubble. If we estimate the transmissions of galaxies in a redshift slice (the green window in the left panel), and plot transmission as a function of galaxy position in one direction on the plane of the sky, we will get a bell-shaped transmission profile: the peak of transmission corresponds to galaxies in the center of the bubble, and the bubble edges are where the transmission drop to almost zero. When measuring the transmission profile through a small window (the orange windows in the left panel), the level of symmetry of the transmission profile around the window center indicates the position of bubbles. 

If the transmission profile on the right and left of the window center are both low and flat (window A in the right panel of Figure~\ref{fig:toyasym}), the window is outside a bubble. If the transmission profiles on both sides of a window are high and symmetric (window B in the right panel), the window is inside a bubble and close to the bubble center. If one side of a window has a flat and low transmission profile, and the other side has a rising transmission profile (window C in the right panel), the window is at the edge of a bubble. By defining and measuring an asymmetry score for these transmission profiles in windows at different positions in a field, it should be easy to detect bubble edges by identifying high asymmetry score regions.

We define an asymmetry vector for each position, $\mathbf{r}$, in an observed field within a circular window using 
\begin{equation}\label{eq:asymvec}
    \mathbf{A}(\mathbf{r})=\frac{1}{N_\mathrm{gal, window}}\sum_{i=1}^{N_\mathrm{gal, window}}{\frac{\mathcal{T}_{i}\cdot\mathbf{r_{i}}}{|\mathbf{r_{i}}|^{2}}},
\end{equation}
where $N_\mathrm{gal, window}$ is the number of galaxies in the window centered at $\mathbf{r}$, $\mathcal{T}_{i}$ is the \lya transmission of $i$-th galaxy, and $\mathbf{r_{i}}=(\mathbf{r_\mathrm{gal,i}}-\mathbf{r})/R_{\rm window}$ is the position of the $i$-th galaxy with respect to the window center, normalized by the window radius ($R_{\rm window}$). In practice, we calculate the asymmetry vector along the horizontal and vertical transverse directions. Looking at the center position of window C in the right panel of Figure~\ref{fig:toyasym}, we see how Equation~\ref{eq:asymvec} reflects asymmetry along the $x$-axis. At the edge of the bubble, only galaxies in the $+x$ direction have high transmission $\mathcal{T}_{i}$, thus the asymmetry vector will be positive with large magnitude. Moving into the center of the bubble, galaxies have high transmission equally in the $-x$ and $+x$ directions, thus $\mathbf{A}(x) \rightarrow 0$. We define the asymmetry score as the magnitude of the asymmetry vector ($|\mathbf{A}|$). 

Figure~\ref{fig:exampasymmap} shows an example ionization map (left panel) and the asymmetry score map (middle panel) generated using the \lya transmission of galaxies in that field (assuming perfect recovery of transmission). In addition, we show the asymmetry vector map (right panel). The asymmetry score map shows that at bubble edges, we recover high asymmetry scores as required. Asymmetry scores in bubble centers and neutral regions are low because of the flat transmission profile. The directions of asymmetry vectors point toward bubble edges. In this paper, for simplicity, we only use the asymmetry score to recover bubble maps. We do not use the direction information of the asymmetry vector in this work, but note that it could be used to further refine bubble finding algorithms.

Figure~\ref{fig:exampasymmap} shows that high asymmetry score areas match real bubble edges. However, to define bubble edges we must calibrate a threshold in asymmetry. To do this, we convert both the input ionization map and asymmetry map into binary maps (neutral/ionized), and compare how similar the two maps are. Detailed descriptions of binary map conversion and calibration can be found in Appendix~\ref{sec:binarymap}. We find the best asymmetry score threshold, $A_{\rm thresh}$, to be $A_{\rm thresh}=0.3-0.4$, and that it is independent of the IGM mean neutral fraction \xHI, meaning it is possible to map bubbles using this method irrespective of the surrounding IGM conditions or bubble sizes.

In what follows, we calculate the asymmetry score using the peak value of the \lya transmission posterior (i.e. the maximum likelihood value) estimated for each galaxy, described in Section~\ref{methods:T}.

\subsection{Suggested parameter setup}\label{sec:bestwindowrnd_xHI}

Our method has three free parameters: asymmetry score threshold ($A_{\rm thresh}$), window radius ($R_{\rm window}$), and window depth. Here we first discuss how altering each parameter changes the result, then recommend the choice of each parameter, finding a single set of parameters which describe bubbles well.

\textit{Asymmetry score threshold}, $A_{\rm thresh}$, is used to convert asymmetry maps into binary bubble maps. $A_{\rm thresh}$ affects the segmentation and the size of the recovered bubbles. A lower $A_{\rm thresh}$ leads to bigger bubble sizes and smoother bubble shapes. When $A_{\rm thresh}$ is set too high, the resulting bubbles no longer represent ionized regions, but the regions around the edges of bubbles.  

\textit{Window radius}, $R_{\rm window}$, in our bubble finding algorithm determines the minimum recovered bubble size and the noise on bubble edge recovery, it should be roughly comparable to the minimum bubble radius. The asymmetry score map produced using a smaller $R_{\rm window}$ reflects the ionization fluctuations on a smaller scale, thus it can be good for finding small bubbles. On the other hand, a big bubble may be split into several small bubbles using a small $R_{\rm window}$. Different reionization morphologies may require different optimal $R_{\rm window}$. However, this mostly impacts the recovery of very small bubbles ($\leq10$\,cMpc). To decide on $R_{\rm window}$, we also need to consider the number of galaxies available for calculating the asymmetry score in each window. A minimum $R_{\rm window}$ can be estimated using the surface density of galaxies $\Sigma_{\rm gal}$. The window radius for a desired average number of galaxies per window, $N_{\rm window}$, will be  
\begin{equation}\label{eq: windowsize}
    R_{\rm window}=\sqrt{\frac{N_{\rm window}}{\pi\sum_{\rm gal}}}.
\end{equation}

\textit{Window depth} is the depth of the window in the redshift direction. It affects the minimum recovered bubble size and the number of galaxies in a field. A deeper window depth will include galaxies in a wider redshift range for the asymmetry map calculation, but will bury the signal from small bubbles.

In Appendix~\ref{sec:asymmetrymapmetrix} we test the best $A_{\rm thresh}$, $R_{\rm window}$, and window depth combination for $\xHI=0.4-0.9$ \citep[i.e. before significant overlap of ionized regions,][]{Miralda-Escude2000}. Encouragingly, we find the best $A_{\rm thresh}$ (0.3), window radius (12 cMpc), and window depth (4 cMpc, $\Delta z\sim$0.014 at $z=8$) remain unchanged across the tested \xHI range, meaning our method works generically for a broad range of bubble sizes. We also test the minimum number of galaxies required to calculate the asymmetry score in a window, finding excluding windows with $N_\mathrm{window} < 2$ reduces noise in the recovered maps.

\subsection{Observational requirements for mapping bubbles in the plane of the sky}\label{sec:asymmap_require}

We now consider what observational conditions are required to map ionized bubbles in the plane of the sky. We test the number density of sources sampled, and the maximum \lya EW\footnote{We use \lya EW limit instead of flux sensitivity because in our case the technique is most sensitive to the EW limit, as we need to detect deviations from the EW distribution for the ionized IGM (Section~\ref{methods:T}).} limit required to recover bubble edges. Full details of our tests are described in Appendix~\ref{app:obs_asym} and we summarize the key results here. We find we need to observe $\simgt0.004$\,galaxies/cMpc$^3$ (corresponding to \MUV$\leq$-17.2 at $z=8$) and a 5$\sigma$ \lya equivalent limit $\leq$ 30\AA\ for the faintest galaxies to obtain the best recovery of our simulated bubble maps: our algorithm can recover bubbles with $R\simgt10$\,cMpc with $<30\%$ bubble size precision in this case, but that we can still trace bubbles well with slightly lower number density ($\simgt0.002$\,galaxies/cMpc$^3$, corresponding to \MUV$\leq$-17.8 at $z=8$). 

To fully map bubbles we need the field of view to be larger than bubble diameters. Typical bubble sizes depend on the reionization timeline and ionizing sources, which we discuss further in Section~\ref{sec:disc_mapping}, but are expected to be $R\sim2-30$\,cMpc ($R\sim2-20$\, arcmin) during the first half of reionization \citep{Lu2024}. To accurately find bubble edges the field of view must also be larger than our window radius (12\,cMpc $\approx$ 5 arcmin at $z=8$). These areas are feasible with wide JWST surveys, particularly in the early stages of reionization.

Figure~\ref{fig:asymmap_2cases_3xHI} shows the asymmetry maps made using \lya transmission in our $z=8$ simulations at $\xHI=0.5-0.9$.
We show the maps obtained using galaxies with $\approx0.002$ galaxies/cMpc$^3$ (ccorresponding to \MUV$\leq$-17.8 at $z=8$) and \lya flux limit corresponding to a 5$\sigma$ \lya equivalent limit $\leq$ 30\AA\ for the faintest galaxies in the map (i.e. brighter galaxies have lower EW limits). We generate the maps by first using the Bayesian method in Section~\ref{methods:T} to infer $p(\mathcal{T})$ for each galaxy given their observed EW, the uncertainty of EW, and \MUV, then calculating asymmetry score at each sky position using the maximum likelihood of $\mathcal{T}$ of galaxies around the position\footnote{We use the maximum likelihood of $\mathcal{T}$ rather than sampling from the inferred P($\mathcal{T}$) distributions because when the EW limit is shallow, P($\mathcal{T}$) is very broad, which we found leads to many false bubbles in the recovered maps. Using the maximum likelihood $\mathcal{T}$ means that galaxies where the transmission is well constrained to $\mathcal{T}\approx0$ or $\mathcal{T}\approx1$ contribute most to the asymmetry score, improving the recovery of the bubbles. This is further demonstrated in Appendix~\ref{app:obs_EW}.}. We set the asymmetry score to 0 for positions with fewer than 2 galaxies in the window to reduce noise in the recovered map.
 
In addition to the number density and flux limit requirements, we also require precise redshift resolution. The window depth must be comparable to the diameter of typical bubbles -- in the first half of reionization the median bubbles have radii of $\simlt10-30$\,cMpc \citep{Lu2024}. Thus, to capture these bubbles, asymmetry maps should be made using galaxies in a thin redshift window (window depth = 4cMpc or $\Delta z\sim0.015$ at $z=8$, as described above in Section~\ref{sec:bestwindowrnd_xHI}), requiring spectroscopic observations \citep[i.e. narrow-bands are too broad to cover single bubbles during most of reionization, see also][for similar conclusions]{Sobacchi2015}.

Finally, to map entire ionized bubbles in the plane of the sky requires a field of view (FoV) that is larger than bubble diameters. For reference, the 1\% largest ionized bubble diameter at \xHI=0.5 (0.9) is $\sim300$ (60)\,cMpc or $\sim$100 (20) arcmin for reionization driven by high mass halos \citep[see Figure~\ref{fig:Rmean}, also][]{Lu2024}. 
In the earliest stages of reionization ($\xHI \simgt0.8$), it should thus be feasible to map entire ionized regions with deep JWST observations. 

Upcoming wide-area imaging and spectroscopy surveys, such as the Euclid Deep survey \citep{Euclid_wide2022, Euclid_deepz6_2022, Marchetti2017b}, Roman High Latitude Spectroscopy Survey \citep[HLSS][]{Wang2022} and Subaru PFS \citep{Greene2022_PFS}, will observe thousands of reionization-era galaxies on $>1$ sq. deg scales required to cover multiple ionized regions in the mid-stages of reionization and observe their \lya. 
However, these surveys will not be deep enough to recover bubble maps by themselves, using the method we describe here, as they will not reach the required galaxy number density and \lya EW limits. However the bright \lya emitters detected in these surveys should signpost large ionized regions and would provide ideal targets for deep spectroscopic follow-up with JWST. We will discuss observational prospects for mapping ionized regions in more detail in Section~\ref{sec:disc_mapping}.

\section{Mapping bubbles along the line-of-sight}\label{sec:LOSbubble}

\begin{figure*}
    \centering
    \includegraphics[width=0.85\textwidth]{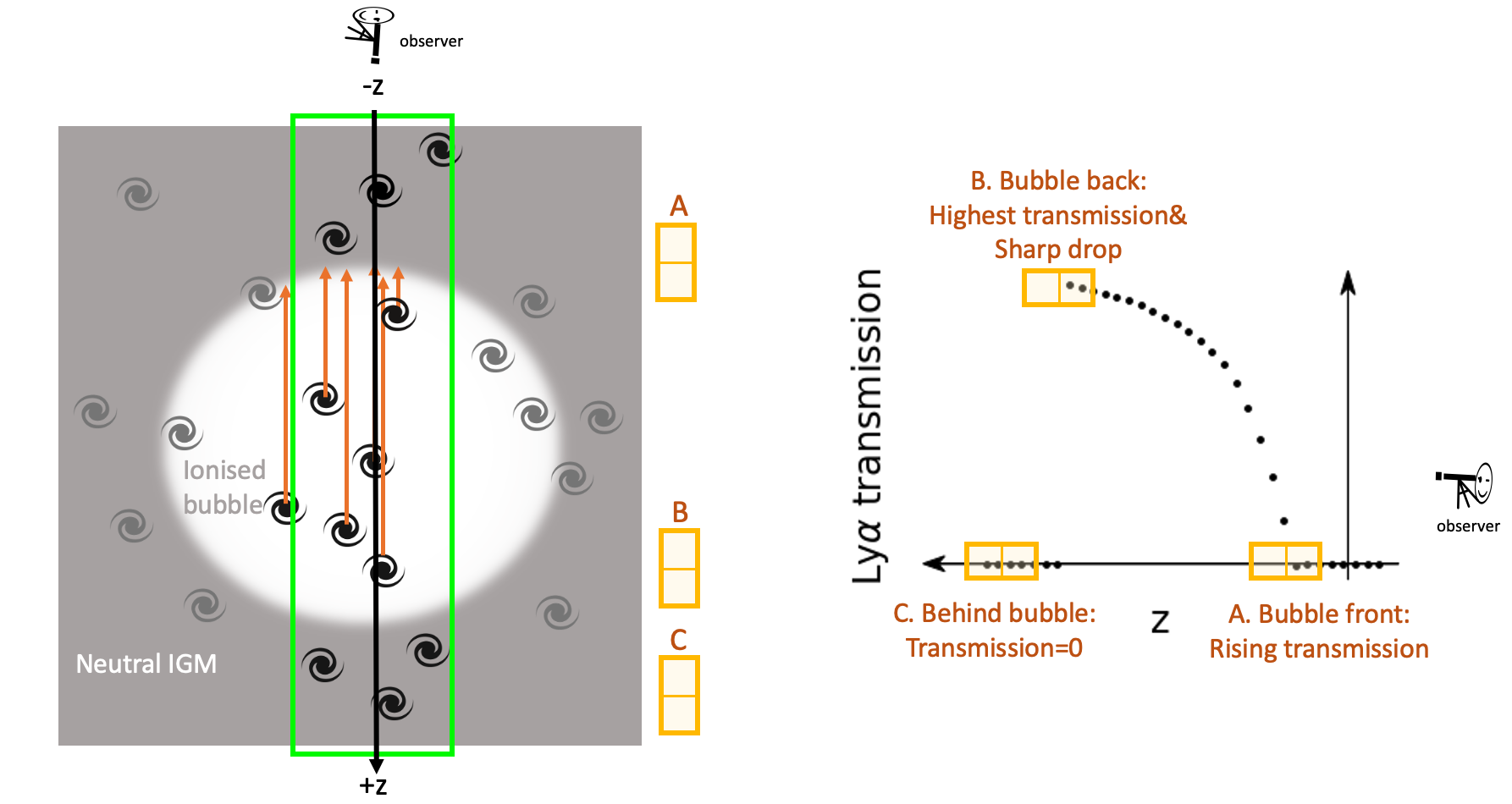}
    \caption{A cartoon to illustrate the transmission profile of galaxies along the line-of-sight (LOS) direction around a bubble. \textit{Left:} Galaxies in and around a spherical bubble. \textit{Right:} The integrated \lya transmissions as a function of LOS distance ($d$) measured using galaxies in the green box in the left panel. Transmission profiles $\mathcal(T)(z)$ viewed in different windows (A, B, and C) are different. \textbf{Window A:} Galaxies in front of the bubble are in fully neutral gas and thus experience a strong \lya damping wing and have almost zero transmission. Stepping back into the bubble the transmission increases because of the increased path length of ionized IGM in front of galaxies. \textbf{Window B:} Galaxies located at the back of the bubble have the highest transmission because they have the longest path length through the ionized IGM. \textbf{Window C:} Galaxies on the far side of the bubble are in fully neutral IGM and have almost zero transmission. Galaxies' integrated \lya transmission is tightly correlated with the distance between the galaxy and the nearest neutral IGM patch along the line-of-sight. We can use this correlation to map ionized regions along the line-of-sight (Section~\ref{sec:LOSbubble}).   }
    \label{fig:cartoon_xz}
\end{figure*}

\begin{figure}
    \centering
    \includegraphics[width=0.88\columnwidth]{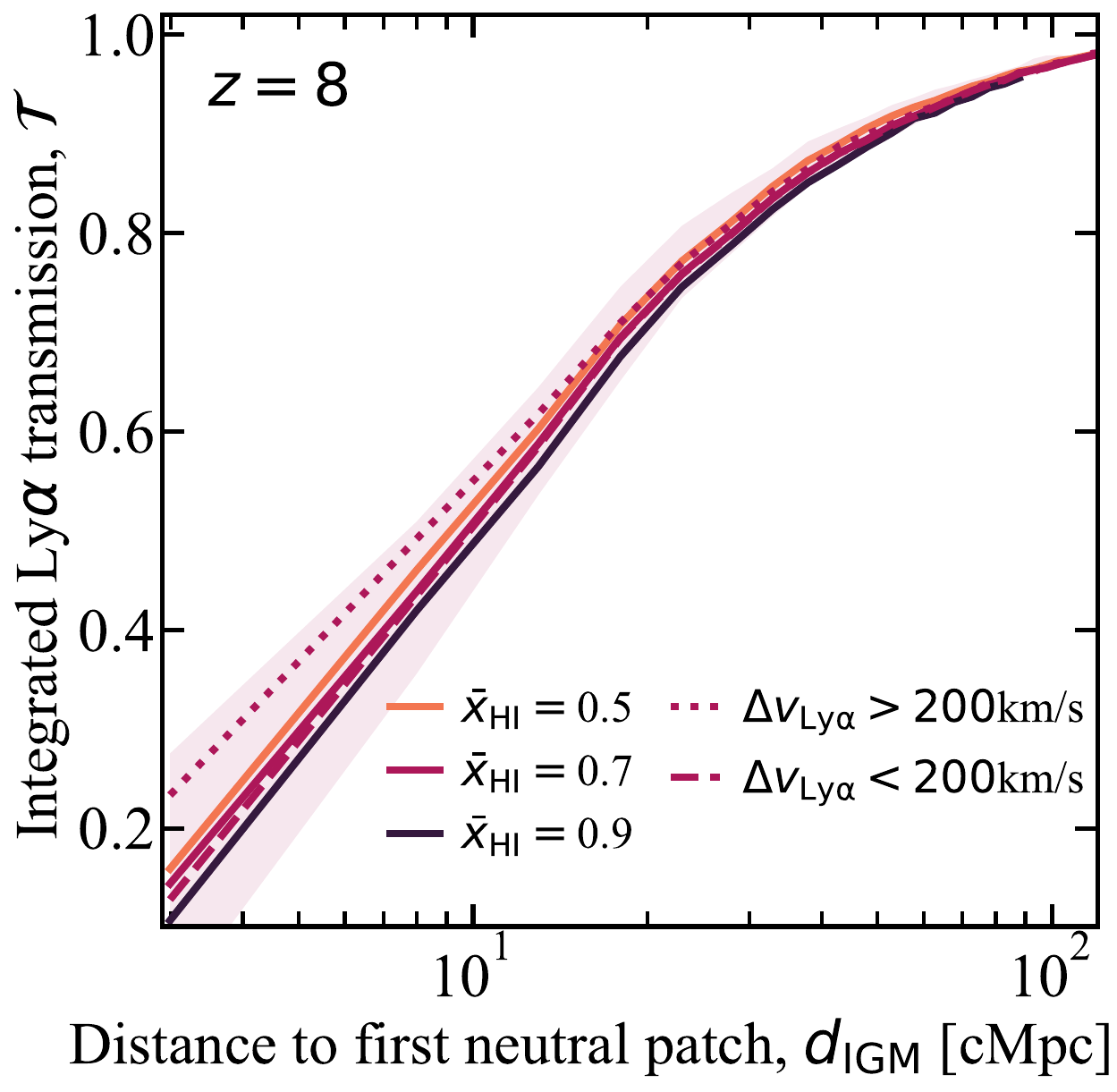}
    \caption{Integrated \lya transmission as a function of the distance between a galaxy and the first neutral IGM patch the galaxy light encounters at $z=8$. Solid lines in orange, maroon, and black curves show $\mathcal{T}(d_\textsc{igm})$ for all the galaxies at \xHI=0.5, 0.7, 0.9, respectively. The shaded region shows the 16-84 percentile range of the \xHI=0.7 curve. Maroon dashed and dotted lines show $\mathcal{T}(d_\textsc{igm})$ at \xHI=0.7, made using galaxies with emitted \lya velocity offsets $\Delta v_{\rm Ly\alpha}>200$km/s and $\Delta v_{\rm Ly\alpha}<200$km/s respectively. None of these curves differ significantly, meaning $\mathcal{T}(d_\textsc{igm})$ does not depend strongly on \xHI or $\Delta v_{\rm Ly\alpha}$, especially for $d_\textsc{igm}>10$\,cMpc. The curves are for $z=8$ only, but the qualitative trends hold at other $z$.}
    \label{fig:T_R}
\end{figure}

\begin{figure*}
    \centering
    \includegraphics[width=0.8\textwidth]{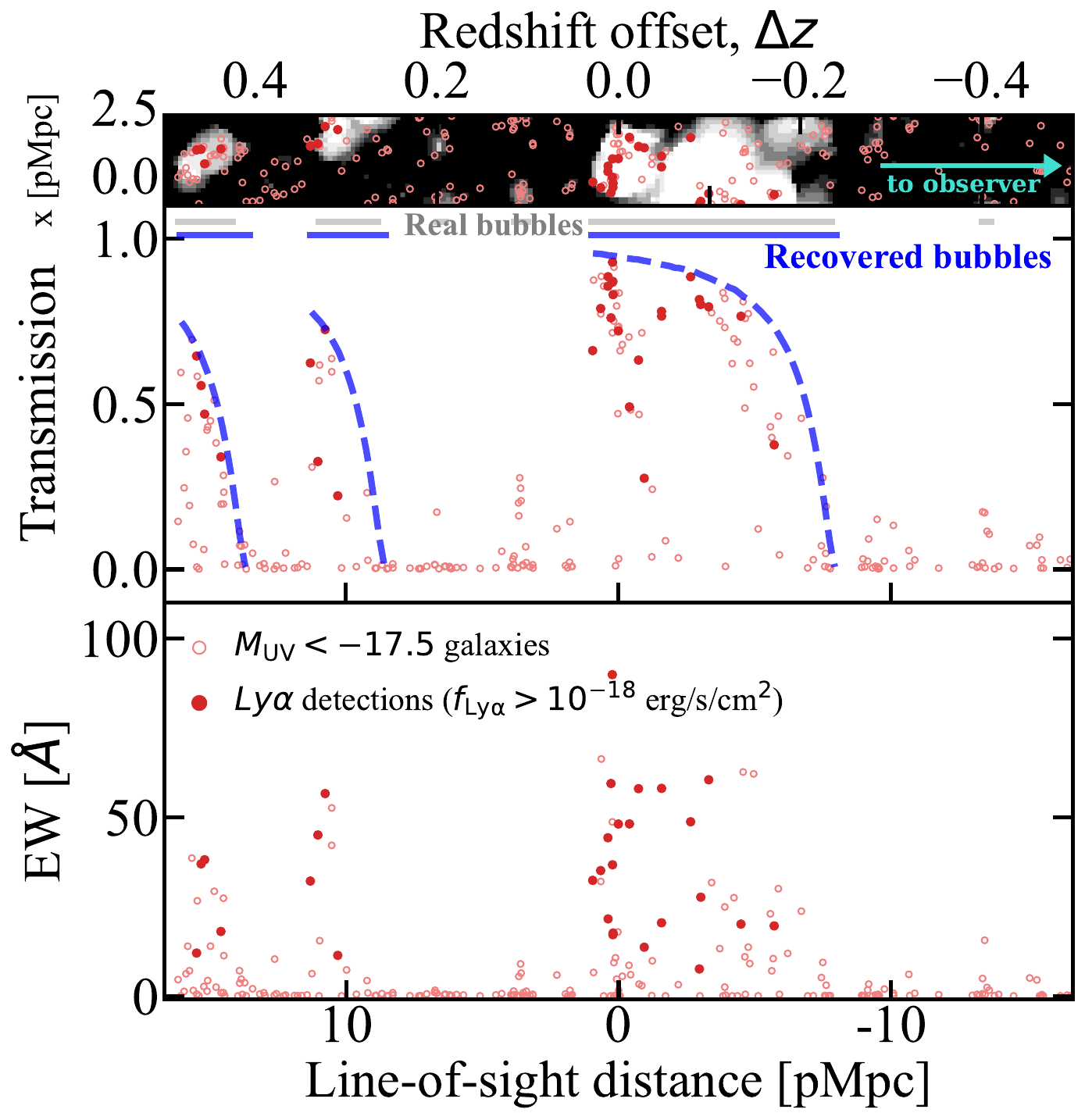}
    \caption{Example sightline skewer showing galaxies' \lya transmissions and EW as a function of line-of-sight distance. Top panel shows the ionization map in the x-z plane (white ionized, black neutral gas). Red empty and solid dots are galaxies with \MUV$>-17.5$ in the slice. Red solid points are galaxies with $f_{\rm Ly\alpha}>10^{-18}$erg/s/cm$^2$. The middle panel shows the \lya transmission for each galaxy as a function of LOS distance. Galaxies in bubbles show high transmission while galaxies in neutral IGM having transmission $\approx0$. Grey horizontal lines show the spans of the simulated ionized bubbles along the sightline. Blue dashed lines are the $\mathcal{T}(d_\textsc{igm})$ curves derived using the damping wing optical depth (Figure~\ref{fig:T_R}), galaxy transmissions follow this curve very well. The blue horizontal lines show the span of the recovered bubbles using our method (Section~\ref{sec:LOS_bub_alg}). The bottom panel shows the observed \lya EW of the galaxies, highlighting that strong \lya EW $\simgt 50$\,\AA\ likely signpost large ionized regions.}
    \label{fig:trans_xz_bubblei63}
\end{figure*}

\begin{figure*}
    \centering
    \includegraphics[width=\textwidth]{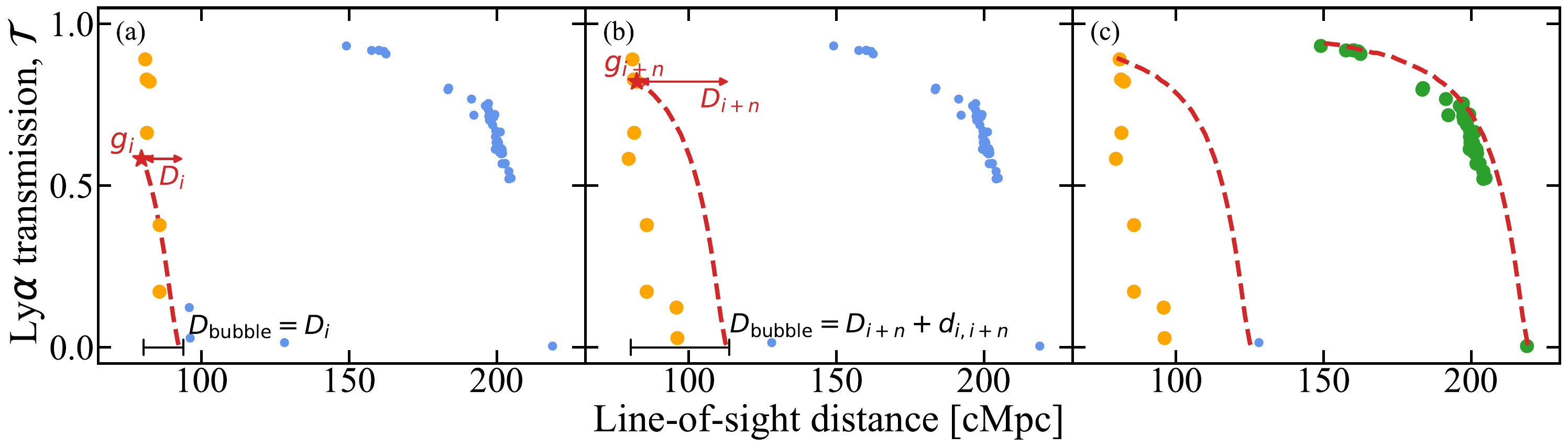}
    \caption{Demonstration of our LOS bubble finding algorithm. 
    \textbf{(a)} Starting from the galaxy farthest away from the observer, $g_{i}$, we invert $\mathcal{T}(d_\textsc{igm})$ (Figure~\ref{fig:T_R}, the red dashed line) to estimate the distance between the galaxy and the nearest neutral IGM patch, $D_i$. We set the LOS bubble diameter, $D_{\rm bubble}= D_i$, and the redshift of the back of the bubble, $z_{\rm back}=z_i$, i.e. the redshift of the first galaxy. We then mark all galaxies in front of $g_i$ within $D_{\rm bubble}$ (orange dots) to be inside the same bubble as $g_i$. 
    \textbf{(b)} We then step through the bubble towards the observer to update our estimate of $D_{\rm bubble}$. For a galaxy $g_{i+n}$ which is a distance $d_{i,i+n}$ from galaxy $g_i$ we estimate the distance of galaxy $g_{i+n}$ from neutral IGM, $D_{\rm i+n}$. If $D_{i+n} + d_{i,i+n} > D_{\rm bubble}$ we update $D_{\rm bubble} \rightarrow D_{i+n} + d_{i,i+n}$. 
    \textbf{(c)} Once we have iterated through all galaxies contained within $D_{\rm bubble}$ and $D_{\rm bubble}$ converges, we continue iterating through all the galaxies in the line-of-sight in the same manner to find all the bubbles. In this example sightline we find two bubbles and their associated galaxies (orange galaxies on the left and green galaxies on the right, the remaining blue point is a galaxy between the two bubbles and is not considered to be inside a bubble). } 
    \label{fig:xz_alg_flowchart}

\end{figure*}

\begin{figure}
    \centering
    \includegraphics[width=\columnwidth]{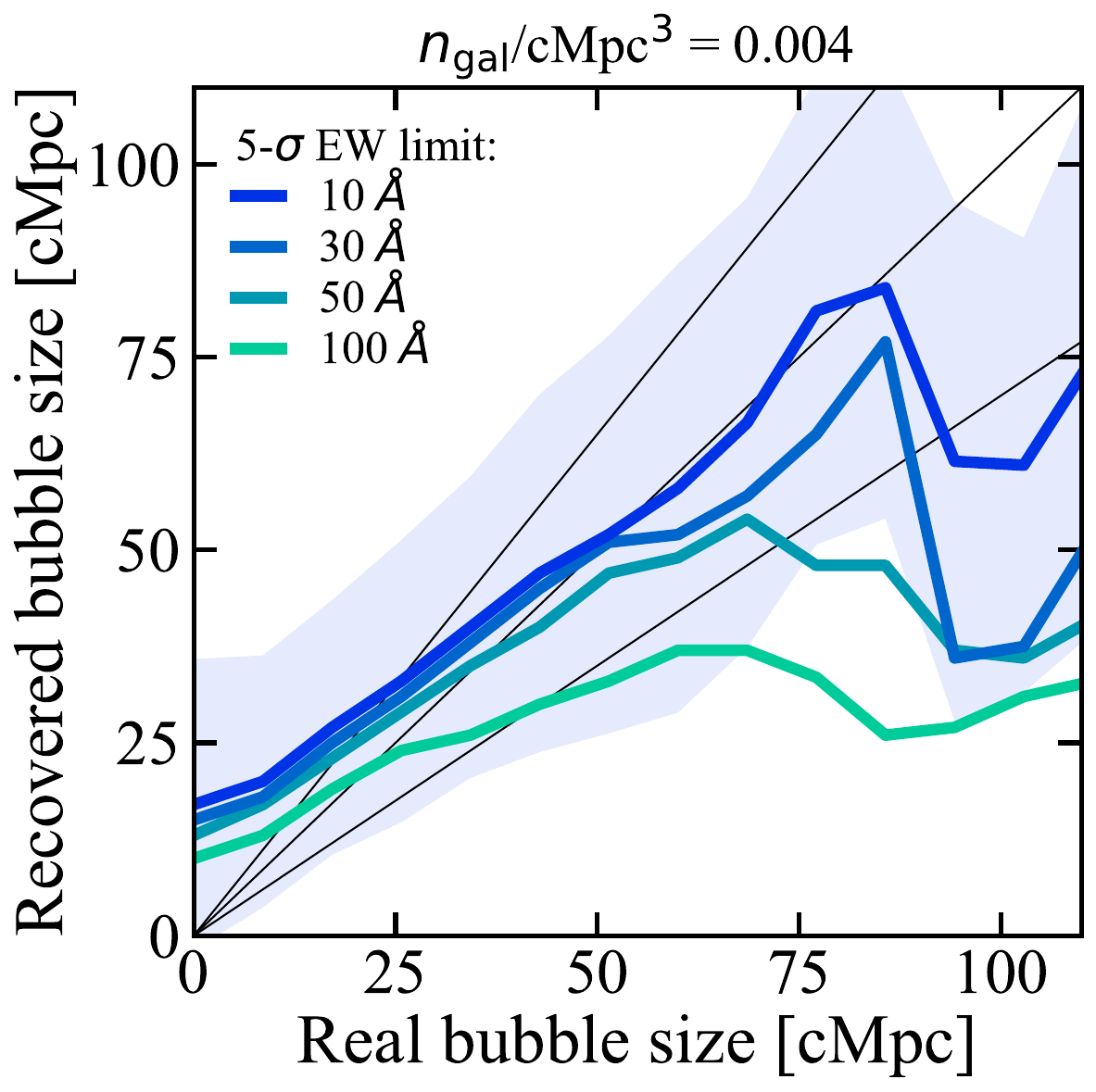}
    \caption{Median recovered bubble size versus real bubble size for LOS bubbles assuming a number density of 0.004 galaxies/cMpc$^3$ (corresponding to complete spectroscopic sampling of galaxies down to $M_{\rm UV, limit}=-17.2$ at $z\approx8$) for 5$\sigma$ \lya equivalent width limits=[10, 30, 50, 100] \AA\ for the faintest galaxies (coloured lines). The shaded region shows the 68$\%$ range of recovered bubble sizes for the \lya equivalent width limit=10\AA\ line. When the \lya equivalent width limit is deeper than 50\AA\ we can estimate $\mathcal{T}$ relatively well for each galaxy and thus we can recover bubble sizes $\sim10-80$\,cMpc to within 30$\%$ of their true values (1-1 and $1-1\pm0.3$ relations marked by solid grey lines). The dip at $R_{\rm real}\sim100$\, cMpc is due to sightlines that pass through a big bubble but without enough galaxies to sample the space, causing us to recover a few smaller bubbles instead of one big bubble. This could be improved in future work when $\mathcal{T}$ can be more precisely estimated, e.g. with a prior from the \lya escape fraction (see Section~\ref{methods:T}). }
    \label{fig:Rrec_Rreal_LOS}
\end{figure}

\begin{figure}
    \centering
    \includegraphics[width=\columnwidth]{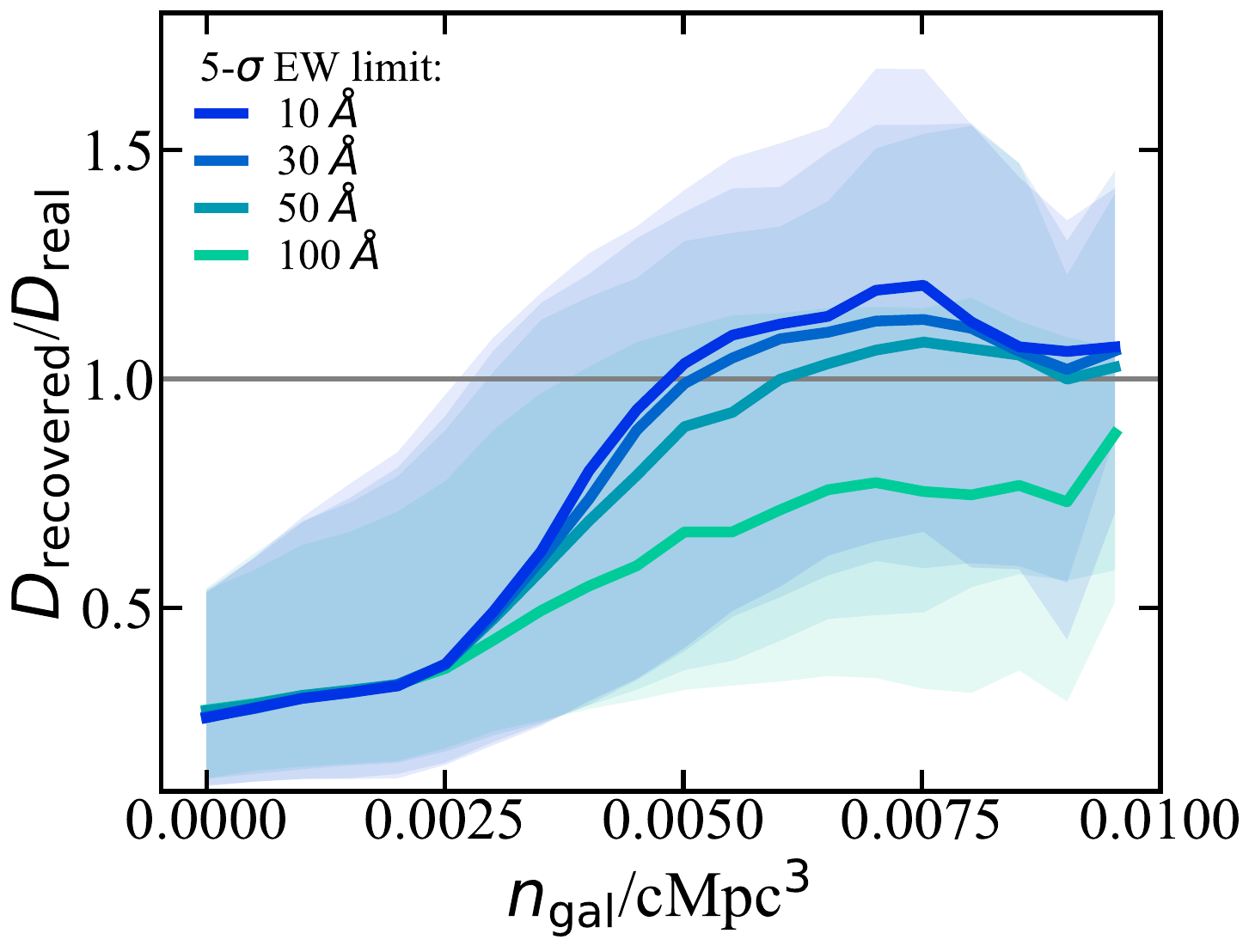} 
    \caption{Median recovered bubble size to real bubble size ratio ($D_{\rm recovered}/D_{\rm real}$) versus galaxy number density for 5$\sigma$ \lya equivalent limits=[10, 30, 50, 100] \AA\, (solid lines) binned by galaxy number density. Shaded regions show the 16-84 percentile of the recovered size ratio. The method converges to the correct bubble size as we reach $\simgt0.004$ galaxies/cMpc$^3$. }
    \label{fig:Rratio_Ngal}
\end{figure}

JWST's field of view (spanning $\approx10$\,cMpc at $z\sim8$) can capture entire bubbles in the early stages of reionization, but JWST can also efficiently map ionized structures along the line-of-sight (LOS) direction at all stages of reionization. The median bubble size at the mid-point of reionization spans $\Delta z \sim 0.1-0.3$. JWST can detect galaxies over such a wide redshift window with precise redshift resolution \citep[R$\sim$1000 corresponding to $\Delta z\lesssim$0.001 with NIRSpec medium resolution grating and NIRCam grism spectroscopy, e.g.][]{Tang2023,Oesch2023,Meyer2024}.
Here we describe a novel algorithm to recover bubble maps in redshift skewers using the LOS transmission profile.

Instead of the bell-shaped \lya transmission profile for bubbles in the plane of the sky (Figure~\ref{fig:toyasym}), the transmission profile in the LOS direction has a shark-fin-like shape (see Figure~\ref{fig:cartoon_xz} for a cartoon illustration). As we move along the LOS away from the observer, the \lya transmission changes as follows: in front of a bubble (closest to the observer), we expect almost no transmitted \lya flux from galaxies. Stepping into the bubble, the closer a galaxy is to the back end of the bubble, the higher the \lya transmission, because of the increasing path length of ionized IGM in front of galaxy. Galaxies at the back of the bubble have the highest \lya transmission. When stepping out of the back size of the bubble, we again expect almost no transmitted \lya flux from galaxies. Thus, the points of non-zero \lya transmission should mark the edges of the bubble.

There is a tight correlation between a galaxy's integrated \lya transmission $\mathcal{T}$ and the line-of-sight distance from the galaxy to the first neutral patch, $d_\textsc{igm}$, which we can use to map the bubble. This is because the damping wing optical depth is dominated by the first neutral IGM along the line-of-sight \citep{Mesinger2008}.  
We derive $\mathcal{T}(d_\textsc{igm})$ and its 16-84 percentile range numerically from our simulations by measuring the $\mathcal{T}$ of galaxies with the same $d_\textsc{igm}$ along $10^5$ sightlines (Section~\ref{sec:21cmfast}). Figure~\ref{fig:T_R} shows $\mathcal{T}(d_\textsc{igm})$ in our simulation at $z\approx8$ for $\xHI=0.5, 0.7, 0.9$. We find $\mathcal{T}(d_\textsc{igm})$ has negligible dependence on \xHI at fixed redshift. We also compare $\mathcal{T}(d_\textsc{igm})$ for galaxies with emitted \lya velocity offset $\Delta v_{\rm Ly\alpha}>200$km/s and $\Delta v_{\rm Ly\alpha}<200$km/s, finding $\mathcal{T}(d_\textsc{igm})$ has only a small dependence on velocity offset once a galaxy is at least $>10\,$cMpc from a neutral patch \citep[see also, e.g.][]{Endsley2022_overdense_bubble,Prieto-Lyon2023b}. We see the variance (coming from the distribution of additional neutral patches along sightlines) of the $\mathcal{T}(d_\textsc{igm})$ curve is also small, resulting in less than 10 cMpc under-/over-estimate of $d_\textsc{igm}$ when estimating $d_\textsc{igm}$ from $\mathcal{T}$.

We note that the relation shown in Figure~\ref{fig:T_R} is only valid for $z=8$, as the damping wing optical depth increases with increasing redshift due to increasing gas density \citep{Gunn1965b}, but the minimal dependence on \xHI and velocity offset for $>10$\,cMpc bubbles is qualitatively true at other redshifts. 
  
Figure~\ref{fig:trans_xz_bubblei63} shows a sightline in our simulation at $\xHI = 0.8$, selected in a sky area corresponding to 3 JWST NIRSpec footprints (42.8 arcmin$^2$ at $z=8$), collapsed along the $y$ direction. We show ionized regions and the positions of all $\MUV < -17.5$ galaxies. We also plot the integrated \lya transmissions and \lya EW of these galaxies as a function of LOS distance, along with our averaged $\mathcal{T}(d_\textsc{igm})$ profile based on the positions and sizes of the recovered bubbles in the skewer. The \lya transmissions of galaxies in each bubble follow our averaged profile very well, with all \lya transmissions falling along or below the curve. 

The scatter of transmission below the $\mathcal{T}(d_\textsc{igm})$ curve mainly comes from bubbles which are clumpy along the line of sight, so galaxies at the same redshift within the field of view are not necessarily at the same distance to neutral gas.
A minor cause of the scatter of transmissions ($\sigma_{\mathcal{T}}<0.05$) is the \lya emission line velocity offset. A strongly red-shifted \lya emission line (e.g. $>300$\,km/s) will suffer less from the damping wing absorption and therefore have higher transmission than the expected value in our $\mathcal{T}(d_\textsc{igm})$ relation.

In this paper we focus on estimating the maximum bubble diameter along the LOS. Therefore we want to find a $\mathcal{T}(d_\textsc{igm})$ curve for each bubble which envelopes the observed transmission spikes, and thus marks the front and back of the bubble. In the next section we describe a simple algorithm to do this.

\subsection{An algorithm for finding bubbles along the LOS}\label{sec:LOS_bub_alg}

To detect bubbles along a LOS using \lya transmission estimates, we develop a simple algorithm. For each galaxy we require an estimate of the integrated \lya transmission (e.g. via comparing the observed \lya EW to the predicted EW based on $z\sim5-6$ observations, as described in Section~\ref{methods:T}), a precise spectroscopic redshift ($\sigma(z) < 0.015$) and sky coordinates so that we can make a 3D map of galaxy positions. The steps of the algorithm are as follows and illustrated in Figure~\ref{fig:xz_alg_flowchart}:
\begin{enumerate}
    \item Starting from the galaxy farthest away from the observer (galaxy with the highest redshift), $g_{i}$, we invert the $\mathcal{T}(d_\textsc{igm})$ curve (Figure~\ref{fig:T_R}) to estimate the distance from $g_{i}$ to the front of the bubble, $D_i$ using the estimated \lya transmission of $g_{\rm i}$.
    \item If $g_{\rm i}$ is the first galaxy in a bubble, we set the redshift of the back of the bubble to the redshift of $g_{i}$, $z_{\rm back}=z_i$.  We set the LOS bubble diameter $D_{\rm bubble}= D_i$, and mark all the galaxies within a distance $D_{\rm bubble}$ in front of $g_{i}$ to be in the same bubble as $g_{i}$ (Panel (a) of Figure~\ref{fig:xz_alg_flowchart}).
    \item A galaxy's \lya transmission may not follow exactly $\mathcal{T}(d_\textsc{igm})$ due to an irregular bubble shape or over-predicted emergent \lya line, and thus may fall below the $\mathcal{T}(d_\textsc{igm})$ line. To make sure all the galaxies in the bubble have transmission falling on or below the $\mathcal{T}(d_\textsc{igm})$ curve, we move to the next galaxy (galaxy with the second highest redshift), $g_{i+1}$, a distance $d_{i,i+1}$ from $g_i$. We estimate the distance of $g_{i+1}$ from the neutral IGM, $D_{i+1}$ by inverting $\mathcal{T}(d_\textsc{igm})$ as above.
    \item If $D_{i+1} + d_{i,i+1} > D_{\rm bubble}$, we update $D_{\rm bubble} \rightarrow D_{i+n} + d_{i,i+n}$ and again mark galaxies within the updated bubble (panel (b) of Figure~\ref{fig:xz_alg_flowchart}).
    \item By iterating through all the galaxies in the sightline, we can identify all the ionized bubbles (panel (c) of Figure~\ref{fig:xz_alg_flowchart}).
\end{enumerate}

To account for the uncertainty in \lya transmission estimates we use the following approach. For each mock galaxy, we use the lower 16\% bound on transmission inferred using the Bayesian approach described in Section~\ref{methods:T} as the input $\mathcal{T}$.
We use this approach rather than the full inferred $P(\mathcal{T})$ or maximum likelihood $\mathcal{T}$ because of the exponential-like relation between distance to HI and \lya transmission (Figure~\ref{fig:T_R}), this would result in a bias to overestimated bubble sizes in our algorithm.
We find using the lower bounds of $\mathcal{T}$ always provides lower bound on bubble size that converges to the true bubble size when we include sufficient galaxies in the sightline (see Section~\ref{sec:LOSbubble_require}).

\subsection{Observational requirements for mapping bubbles along the line-of-sight}\label{sec:LOSbubble_require}

To detect bubbles along the line-of-sight requires (1) a good estimate of \lya transmission for each galaxy, which depends on the \lya EW limit, and (2) a sufficient number of galaxies in the line-of-sight to sample the region. To test the required observational constraints we apply our algorithm to sightlines extracted from our simulations, assuming different spectroscopic completeness and \lya EW limits.

We sample 900 $10\,\mathrm{cMpc}\times10\,\mathrm{cMpc}\times300\,\mathrm{cMpc}$ sightlines from our $\xHI=0.8$ simulation cube, corresponding to individual JWST NIRSpec pointings spanning $\Delta z \approx 1$ at $z\sim 8$. For each galaxy in each sightline, we infer $p(\mathcal{T})$ using the Bayesian approach in Section~\ref{methods:T}. 
We apply our algorithm to detect bubbles along these sightlines given galaxy positions and the lower limit of the inferred $p(\mathcal{T})$ (see discussion in Section~\ref{sec:LOS_bub_alg} above).

In Figure~\ref{fig:Rrec_Rreal_LOS} we plot the median recovered bubble diameter, $D_\mathrm{recover}$, from our algorithm versus true bubble diameter, $D_\mathrm{true}$, for $\sim$5000 bubbles.
We calculate the true bubble positions and diameters in each sightline by collapsing the IGM skewer in both spatial directions, measuring the mean neutral fraction at each redshift position in the 1D skewer, and classify regions that have mean neutral fraction below $x_{\rm HI}=0.5$ as ionized. We match the recovered bubbles to the true bubbles by iterating over the true bubbles and searching for recovered bubbles that overlap with them. If a true bubble has no corresponding recovered bubble, we record the size of the recovered bubble as 0 cMpc. If more than one recovered bubble overlaps with a true bubble, we count both recovered bubbles.
We compare the results assuming a number density $\approx0.004$ galaxies/cMpc$^3$ in each skewer (corresponding to complete spectroscopy of $\MUV\leq-17.2$ galaxies at $z\approx8$ in our simulations) using \lya flux limits corresponding to $5\sigma$ EW limits of 10-80\,\AA\ for the faintest galaxies (i.e. brighter galaxies will have lower EW limits). For $>80$\,cMpc bubbles, the median recovered bubble size drops because some sightlines pass through big ionized bubbles but lack sufficient galaxies to sample the space, resulting in big ionized bubbles being recovered to be spatially close smaller bubbles. We note that this 'bubble splitting' issue can be reduced if the estimation of $\mathcal{T}$ is improved.

We see our algorithm can robustly recover the sizes of bubbles along sightlines extremely well, to $\simlt5$\% for $D_{\rm true}\approx$10-80 cMpc bubbles, providing the flux limit corresponds to a \lya EW limit of $\simlt30\,$\AA, roughly the median EW expected in a fully ionized IGM (see Section~\ref{methods:T}). Even with an EW limit of $\simlt50$\,\AA\ we can obtain a lower limit on the bubble size within 30\% of the true size. For true bubbles with $D_{\rm true}\lesssim$15 cMpc, we recover larger bubble diameters because we set the mean $x_{\rm HI}$ threshold to 0.5 when identifying true bubbles in 1D skewers, which captures only the most ionized region at the center of small bubbles. 

We find the number density of galaxies along a sightline is the dominant factor for the accuracy of the recovered bubble size. Figure~\ref{fig:Rratio_Ngal} shows the ratio between recovered bubble size and real bubble size versus galaxy number density along sightlines, obtained by sampling recovered bubbles in our simulations varying the $M_\mathrm{UV,limit}$ detection limit between -19 and -17 to sample a range of observed galaxy number densities. We find we can recover the true bubble size within $<30\%$ when the observed number density of galaxies reaches $\simgt0.004/$cMpc$^3$ and $D_{\rm recover}/D_{\rm true}$ converges to $\approx1$ when $\simgt0.005/$cMpc$^3$. We recover a slightly larger ($1.1-1.2\times$) median bubble size compared to the `true' bubble size because of the irregular shapes of bubbles, where small ionized projections at the start or end of bubbles are identified via strong \lya emission by our mapping algorithm but can be missed in our calculation of `true' bubble positions in the 1D skewers. Future works carefully accounting for the 3D shapes of bubbles will improve the bubble size comparison.

We also test the effect of the galaxy redshift uncertainty and find it has a minor impact on the recovered bubble size as long as it is smaller than the bubble diameter. We find our method requires a redshift uncertainty of $\sigma_{z}\simlt0.015$ ($\sim5$ cMpc at $z=8$). This requirement is achievable with JWST NIRSpec grating and NIRCam grism spectroscopy (which can measure $\Delta z\approx 0.001$).

\section{Discussion}\label{sec:discussion}

JWST's ability to spectroscopically confirm and detect \lya from $\MUV \simlt -17$ galaxies opens a new window on reionization, providing the potential to map ionized regions in four dimensions for the first time. This is a substantial step beyond the volume-averaged view we have had of the $z>6$ IGM prior to JWST \citep[e.g.,][]{Stark2010,Schenker2014,Pentericci2014,Greig2017,Davies2018b,Mason2018,jung_texas_2020,Bolan2022}.
We have developed ionized bubble mapping methods which, for the first time, use information from ensembles of galaxies (see also our companion paper, Nicolić et al. in prep.). In doing so, our methods mitigate the effects of intrinsic variability of \lya transmission to enable robust estimates of the positions and sizes of ionized bubbles. In this section we discuss optimal strategies for mapping individual ionized regions to understand reionization on a local scale (Section~\ref{sec:disc_mapping}) and prospects for constraining the bubble size distribution (Section~\ref{sec:disc_BSD}).

\subsection{Optimal observational strategies for mapping ionized regions} \label{sec:disc_mapping}

\begin{figure}
    \centering
    \includegraphics[width=\columnwidth]{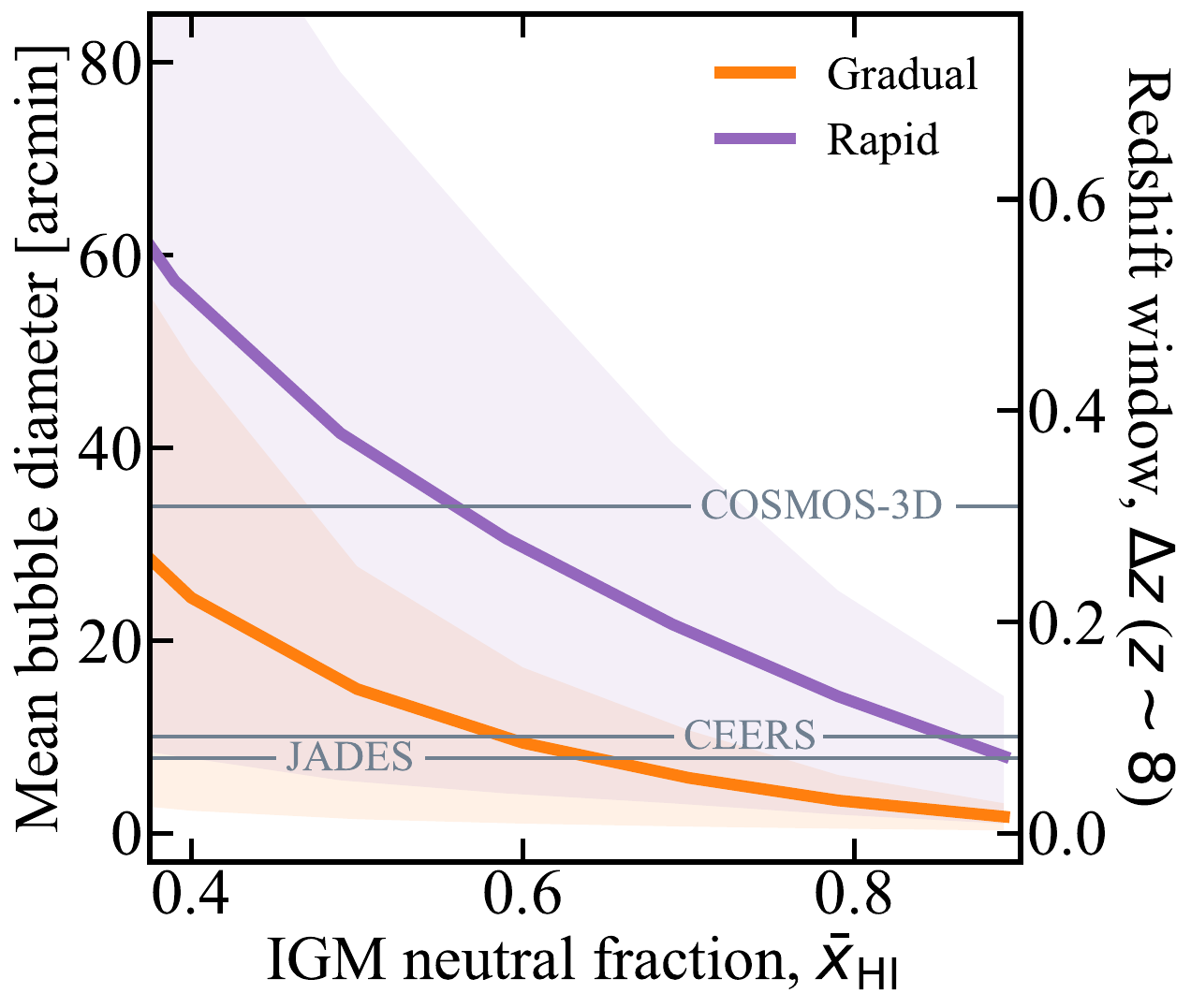}
    \caption{Mean bubble diameter as a function of \xHI. We adopt the characteristic bubble size measured using the mean-free-path method in \citet{Lu2024} and assume $z=8$ (there is miminal impact on the physical sizes over the epoch of reionization, $z\sim6-10$). We show bubble diameter in arcmin on the plane-of-the-sky on the left axis and along the line-of-sight in redshift width on the right axis. Grey lines mark the area of existing and upcoming spectroscopic surveys with JWST.}
    \label{fig:Rmean}
\end{figure}

In Sections~\ref{sec:asymscore} and ~\ref{sec:LOSbubble} we presented two simple methods for mapping ionized regions in the plane of the sky and along the line of sight. To map an ionized region and estimate its size requires observing a volume sufficient to detect the edges of the bubble, a sufficient number density of spectroscopically confirmed galaxies to trace the volume, $\simgt0.002-0.005$ galaxies/cMpc$^3$, systemic redshifts (i.e. from non-resonant lines like [OIII]5007) with high redshift precision $\simlt 0.015$ to map galaxies on scales smaller than the expected bubble sizes, and deep \lya observations, reaching $5\sigma$ rest-frame EW limits of $\approx30$\,\AA\, for the faintest (\MUV$\sim$-17) surveyed galaxies, to estimate the transmission of \lya through the IGM, within reach of deep JWST observations \citep[e.g.][]{Jones2024,Tang2024c}. 

To understand the required volumes for mapping ionized bubbles in Figure~\ref{fig:Rmean} we show the diameter of ionized bubbles (median and 68\% range) predicted by the simulations by \citet{Lu2024}, in both the plane of the sky in arcmin (left axis) and along the line-of-sight in $\Delta z$ (right axis) at $z=8$. We show the predictions for two reionizing source models, a gradual reionization driven by galaxies in low mass ($M_{\rm h}>5\times10^{8}M_{\odot}$) halos, versus a more rapid reionization driven by galaxies in higher mass halos ($M_{\rm h}>10^{11}M_{\odot}$). These models should span the expected range of bubble sizes. We compare the bubble sizes of ionized regions to the field of view of various JWST Cycle 1-3 spectroscopy surveys (assuming square survey areas): JADES GOODS-N + GOODS-S \citep{Bunker2023b}, CEERS \citep{Finkelstein2022} and COSMOS-3D \citep{Kakiichi2024jwstprop}. We note only CEERS and JADES obtained \lya\ spectra. As noted in Section~\ref{sec:asymmap_require}, upcoming wide-area spectroscopic surveys with e.g., Roman, Euclid, PFS, MOONS \citep[][Hartman et al., in prep]{Cirasuolo2012,Euclid_deepz6_2022,Euclid_wide2022,Wang2022,Greene2022_PFS, Maiolino2020_MOONRISE} will span $>1$ sq.~deg and contain many bubbles signposted by bright \lya-emitters, but are not sensitive enough by themselves to detect the high number density of galaxies we need for mapping. Thus deep follow-up with JWST will be essential.

From Figure~\ref{fig:Rmean} it is clear that mapping whole ionized regions in the plane of the sky with JWST becomes most feasible when $\xHI \simgt 0.5$. In the mid-stages of reionization ($\xHI\sim0.4-0.6$) the field of view of the marked JWST surveys may be covered by a single ionized region. This is consistent with the high field-to-field variation in the \lya fraction measured from current $\sim100$ sq. arcmin JWST fields \citep{Napolitano2024,Tang2024c}. Wide area JWST spectroscopy in fields with high \lya detection fractions should be able to map whole bubbles in the mid-to-early stages of reionization. Figure~\ref{fig:Rmean} demonstrates that mapping bubbles along the line of sight is an efficient approach for estimating bubble sizes with JWST at all stages of reionization, as the largest bubbles at $\xHI\simgt0.5$ span $\Delta z <0.7$, which can be easily mapped with the wide wavelength coverage of JWST NIRSpec and NIRCam grism.

In the past few years, ground-based observations have found evidence for clustered \lya emitters at $z\sim7$ \citep[e.g.,][]{Vanzella2011,Tilvi2020,Jung2022}, implying the presence of ionized regions, but have been limited in the ability to detect \lya from UV faint ($\MUV \simgt -20$) galaxies.
Excitingly, early JWST studies have demonstrated that comprehensively mapping bubbles along the line of sight is now finally feasible. \citet{Chen2024}, \citet{Witstok2024} and \citet{Tang2024c} recently showed that with JWST spectroscopy it is possible to start mapping out the 3D distribution of galaxies and their \lya around high EW ($\simgt 50$\,\AA) \lya-emitting galaxies, which likely signpost ionized regions. JWST has vastly increased the number of detectable sources compared to ground-based studies. Our line-of-sight bubble mapping method requires a density of $\simgt 0.004$ galaxies/cMpc$^3$, corresponding to complete spectroscopic observations of $\MUV \simgt -18 (-17.2)$ galaxies at mean density at $z\approx7 (8)$ \citep[e.g.,][]{Bouwens2021}, but these densities could be obtained with a shallower detection limit in an overdense sightline, to obtain an estimate of the expected bubble size. The outlook for identifying and obtaining spectra of these required galaxy number densities is excellent with JWST: for example, around one of the strongest known $z>7$ \lya\ emitters, JADES-13682 \citep[$z=7.3$, EW$\approx400$\,\AA,][]{Saxena2023}, estimates of a factor $\sim7-25\times$ overdensity have been reported using NIRCam imaging and grism spectroscopy \citep{Endsley2023,Witstok2024,Tang2024c}. Reaching EW $\approx 30$\,\AA\, for $\MUV=-18$ ($-17.2$) galaxies requires $\approx8$ ($45$) hrs with JWST NIRSpec \citep[e.g.][]{Jones2024, Tang2024c}, feasible with deep JWST programs. 

The methods we have presented here could be further refined in several ways in future work.
As described in Section~\ref{methods:T}, estimates of \lya\ escape fractions could be used as a prior on $p(\mathcal{T})$, which would further improve our estimates of the IGM transmission and thus bubble maps.
Additionally, with deep high-resolution ($R>2000$) spectroscopy to resolve the \lya lineshape at high S/N, it could be possible infer the wavelength-dependent damping wing attenuation \citep{Mason2020}, rather than just the integrated transmission. This would provide a more precise estimate of \lya transmission for each galaxy and thus of the bubble size and position (Nikolić in prep.). 
Furthermore, in this work we have not considered the impact of self-shielding systems on $<$\,cMpc scales inside ionized regions, which would add additional \lya opacity \citep[e.g.,][]{Bolton2013, Mesinger2015,Mason2020,Nasir2021}. However, these absorbers are predicted to be small ($\approx 1-20$\,ckpc) and not uniformly distributed in the IGM and thus the probability of the majority of galaxies intersecting an absorber along the line of sight is low. We note that \citet{Chen2024} demonstrated that strong small-scale ($< 100$\,pkpc) galaxy overdensities may contain dense neutral gas which attenuates \lya, though this is on much smaller scales than the expected sizes of bubbles. Future systematic observations of \lya visibility as a function of overdensity will be important for understanding systematics in our methods due to residual neutral gas in ionized regions. However, as our methods utilize ensembles of galaxies and rely on `edges' in \lya transmission at fixed redshift, rather than absolute transmission, our approach should not be strongly impacted by the inclusion or redshift evolution of local small-scale absorbers in the ionized IGM or CGM.

The prospects for obtaining larger samples of ionized regions along multiple sightlines are promising. Upcoming wide-area imaging and spectroscopic surveys as described above, and on-going \lya narrow-band surveys \citep[e.g.][]{Ouchi2017,Hu2019} will provide thousands of bright $z\simgt7$ galaxy candidates and \lya-emitters (flux $\simgt 1\times10^{-17}$\,erg\,s$^{-1}$\,cm$^{-2}$).
At $z\simgt7$ these bright galaxies and \lya-emitters are likely to trace overdensities and sit in ionized regions. These regions would be ideal for spectroscopic follow-up with JWST to spectroscopically confirm faint sources ($\MUV \simgt -18$) to map overdensities in 3D and obtain \lya spectroscopy to infer the bubble sizes.
A key prediction of reionization simulations is that reionization starts in overdensities \citep[e.g.,][]{Iliev2007,Mesinger2007,Trac2011a,Ocvirk2016,Hutter2020}, with galaxy properties playing a role in shaping the size distribution of ionized regions \citep[e.g.,][]{McQuinn2007b,Seiler2019a,Lu2024}. Larger samples would enable the first systematic tests of bubble sizes as a function of overdensity and galaxy properties to enable an understanding of reionization on a local scale as we could directly compare bubble sizes to the estimated ionizing emission from the detected galaxies in the region. This could provide new insight in the ionizing photon escape fraction, star formation histories of the visible galaxies, and the contribution of galaxies below JWST's detection limits.

\subsection{Prospects for estimating the size distribution of ionized regions}\label{sec:disc_BSD}
With a sufficiently large survey it should be possible to infer the bubble size \textit{distribution} as a function of redshift. 
This distribution is sensitive to the halo mass scale of the dominant ionizing sources with reionization driven by higher mass halos producing rarer, larger bubbles at fixed \xHI compared to low mass halos, \citep[e.g.,][]{McQuinn2007b,Seiler2019a,Lu2024} and thus offers important insights in the drives of reionization. 


Since mapping bubbles along sightlines does not require wide continuous survey areas, we could efficiently build up a `bubble size distribution' using multiple JWST fields (with spatially separated pointings to avoid having to model correlations). We note this will require appropriate definition of bubble sizes for comparison between models and observations as our line-of-sight method compresses 3D information effectively into 1D skewers where we trace the distribution of ionized `lengths' in the IGM.

To explore how many fields are necessary to estimate the bubble size distribution we perform a Monte Carlo simulation using our IGM simulations. We take $10\,\mathrm{cMpc}\times10\,\mathrm{cMpc}\times300\,\mathrm{cMpc}$ (corresponding to single JWST NIRSpec pointings spanning $\Delta z \approx 1$ at $z\sim 8$) from our simulation cubes, as described in Section~\ref{sec:LOSbubble_require}. We use simulations spanning $\xHI=[0.4-0.9]$ to capture a range of bubble size distributions. We use our sightline bubble mapping technique to estimate the sizes of the ionized regions in each skewer.

\begin{figure}
    \centering
    \includegraphics[width=\columnwidth]{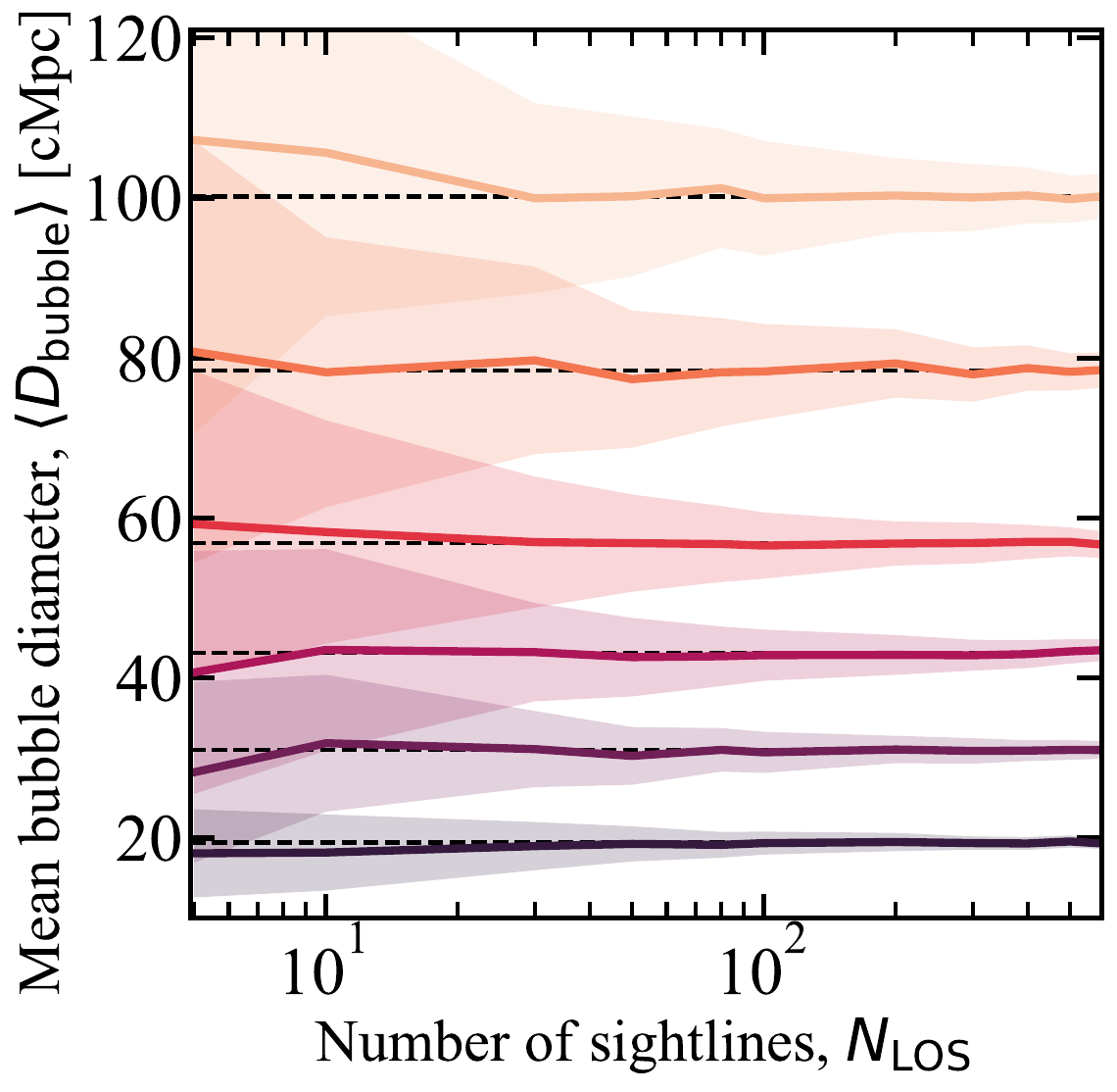}
    \caption{Mean bubble diameter recovered from measuring bubbles along $N_\mathrm{LOS}$ sightlines using our bubble mapping algorithm.(Section~\ref{sec:LOSbubble}), for simulations with different bubble size distributions. The solid and shaded regions show the median and 68\% range of the recovered mean bubble size from 150 realizations. With $\simgt15-30$ sightlines it would be possible to distinguish different bubble size distributions.}
    \label{fig:Rmean_LOS}
\end{figure}

In Figure~\ref{fig:Rmean_LOS} we plot the recovered mean bubble diameter as a function of the number of sightlines used, for simulations with different bubble size distributions, obtained from sampling 150 realizations of our sightlines. We see that we can distinguish differences in mean bubble diameter of $\simgt10$\,cMpc with just $15-30$ independent sightlines. The prospects for building up samples of this number of sightlines is very feasible, for example, JWST pure-parallel NIRCam imaging and slitless spectroscopy surveys are observing $\simgt200$ pointings in Cycles 1-3 which would be ideal for spectroscopic follow-up \citep[e.g.,][]{Williams2021,Morishita2023_beacon,Egami2024}, and \lya spectroscopy has already been obtained in four independent sightlines in Cycles 1-2 \citep{Jung2023,Jones2024,Napolitano2024,Tang2023,Tang2024c}.

\section{Conclusions}\label{sec:conclusions}
JWST provides our first opportunity to chart the reionization process in four dimensions. In this paper, we develop two simple methods to map ionized bubbles using ensembles of galaxies, allowing us to average over variation in intrinsic \lya emission. Our main conclusions are as follows:
\begin{enumerate}
    \item We develop an edge detection algorithm to map ionized bubbles in the plane of the sky, based on the asymmetry of \lya transmission at each position on the sky. Our method is independent of \xHI and can robustly map bubbles $\simgt10$\,cMpc.
        \item Mapping bubbles in the plane of the sky requires a galaxy number density of $\simgt0.002$\, galaxies/cMpc$^3$ \citep[corresponding to $\MUV \simlt -18.5(-17.8)$ at $z\sim7(8)$, e.g.][]{Bouwens2021} to trace the density field (with the most precise recovery for $n_\mathrm{gal}\simgt0.004$/cMpc$^3$), survey areas greater than the typical sizes of ionized regions ($\sim25-3600$\,arcmin$^{2}$ for $\xHI\sim0.9-0.4$), and accurate galaxy redshift determination ($\Delta z\simlt0.015$). We require a 5$\sigma$ \lya equivalent width limit of $\simlt$30\AA\ for the faintest galaxies.
        These sensitivity requirements are not within reach of currently planned wide-area spectroscopic surveys (PFS, MOONS, Euclid-Deep, Roman HLS), however these surveys will find bright \lya-emitters which likely signpost large ionized regions and are thus ideal for deeper follow-up with JWST. Covering the areas required for mapping bubbles will be possible in the early stages of reionization with wide-area JWST spectroscopy.
    \item An efficient strategy to map bubbles at all stages of reionization with JWST is along the line-of-sight, which can be performed even in a single NIRSpec pointing, as bubble diameters are expected to span a redshift range $\Delta z \sim0.05-0.6$ for $\xHI\sim0.9-0.4$. We develop an algorithm which makes use of the nearly one-to-one relation between integrated \lya transmission and distance from a galaxy to the first neutral IGM patch, the relation is calibrated using inhomogeneous large-scale IGM simulations \citep{Lu2024}.
        \item The spectroscopic and imaging capabilities of JWST already provide the sensitivity to
        map bubbles along the line-of-sight. 
        We find we require a number density of $\simgt 0.004$ galaxies/cMpc$^3$ and  EW limit of $\approx 30$\,\AA\ to accurately recover bubbles $\simgt10$\,cMpc, which is attainable with deep spectroscopy of $\MUV \simlt -18(-17.2)$ sources at $z\approx7(8)$, and/or in overdensities of brighter sources. Shallower observations should still provide robust lower limits on bubble sizes.
\end{enumerate}

Early JWST studies have already demonstrated the potential of JWST spectroscopy to map the 3D distribution of galaxies in ionized regions \citep{Chen2024,Witstok2024,Tang2024c}. By constraining the sizes of these ionized regions based on estimates of galaxies' \lya transmission we will be able to link galaxies directly to the growth of ionized regions, enabling a local understanding of reionization. Additional constraints on IGM transmission from \lya escape fractions and line profiles will further improve bubble mapping methods, potentially reducing constraints on galaxy number densities (Nikolić in prep.). Accurate bubble maps will provide the opportunity for ionized photon `accounting': enabling constraints on the escape fraction at $z>6$ and the ionizing contribution of galaxies below JWST's detection limits. Future observations, sampling a range of environments, could start to estimate the size distribution of ionized regions and quantitatively establish how overdensity is linked to the growth of ionized regions.

Our code to create ionized bubble maps in the plane of the sky and along sightlines given galaxy positions, Lya transmissions, and redshifts will be available upon acceptance of this paper.

\begin{acknowledgements}
    We thank Dan Stark for useful discussions. TYL, CAM and AH acknowledge support by the VILLUM FONDEN under grant 37459. CAM acknowledges support from the Carlsberg Foundation under grant CF22-1322. AM acknowledges support from the Italian Ministry of Universities and Research (MUR) through the PRIN project "Optimal inference from radio images of the epoch of reionization", the PNRR project "Centro Nazionale di Ricerca in High Performance Computing, Big Data e Quantum Computing".  The Cosmic Dawn Center (DAWN) is funded by the Danish National Research Foundation under grant DNRF140. This work has been performed using the Danish National Life Science Supercomputing Center, Computerome.

\end{acknowledgements}

%
%
\bibliographystyle{aa}
\bibliography{library}{}

\begin{appendix}
\section{Optimizing bubble map recovery from asymmetry maps}\label{sec:asymmetrymapmetrix}

Here we describe how we calibrate the parameters used for our asymmetry mapping technique (Section~\ref{sec:asymscore}). We describe how we compare the recovered maps to the simulated ionization maps in Appendix~\ref{sec:binarymap}, our comparison metric in Appendix~\ref{sec:bestmetric} and the shape of the window function for creating asymmetry maps in Appendix~\ref{sec:bestwindowrd}.

\subsection{Creating binary \xHI and asymmetry maps}\label{sec:binarymap}
\begin{figure}
    \centering
    \includegraphics[width=\columnwidth]{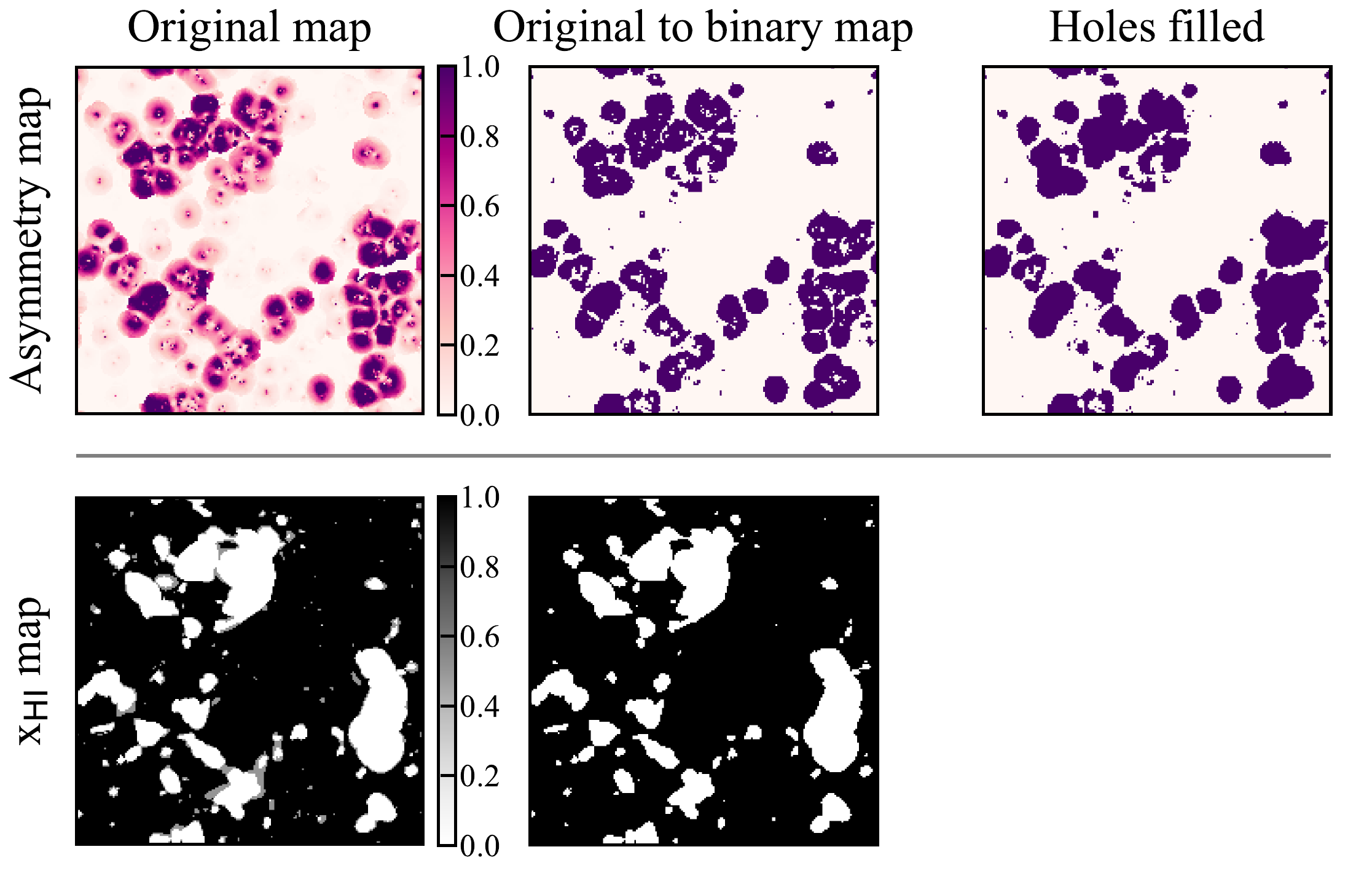}
    \caption{An example of converting an asymmetry score map into a binary ionization map (Top panel), and converting an ionization map into a binary ionization map (Bottom panel). The left column shows the original maps. The second column shows the maps converted into binary maps by setting all the pixels with asymmetry score ( \xHI) above (below) a threshold ionized. The third column shows the binary map after we use $\texttt{scipy.ndimage.binary\_fill\_holes}$, to fill the holes, which correspond to areas inside bubbles, in the second column.}
    \label{fig:binamapdemo}
\end{figure}

The asymmetry map corresponds to the positions of bubble edges, not the positions of ionized regions. To compare the recovered asymmetry maps to the original ionization maps, we need to fill the recovered bubble edges. Here we describe how we convert the asymmetry and the original ionization maps into binary maps, in which 0 (1) in the map means neutral (ionized).

We convert our simulated ionization maps into binary maps to compare with our recovered maps. We first collapse the 3D ionization cubes into 2D binary maps by taking the mean neutral fraction over the line-of-sight direction. We then convert the neutral fraction at each sky position into a binary value by setting the position to be ionized if its mean neutral fraction is less than 0.5. The lower panel in Figure~\ref{fig:binamapdemo} shows the mean neutral fraction for the ionization slice (first column) and the binary ionization map.

To convert the asymmetry maps to binary maps we first set an asymmetry score threshold ($A_{\rm thresh}$). Pixels with asymmetry score above (below) the threshold are marked as ionized (neutral). We find $A_{\rm thresh}\approx0.3-0.4$ provides the best map recovery (see below). In the upper panel of Figure~\ref{fig:binamapdemo}, we show an original asymmetry map in the first column and the binary map in the second column. 
Secondly, since high asymmetry scores reflect bubble edges, we use the $\texttt{scipy.ndimage}$ function, $\texttt{binary\_fill\_holes}$, to fill the closed outlines in the binary asymmetry map. The final binary asymmetry map is shown in the upper right corner in Figure~\ref{fig:binamapdemo}.

\begin{figure*}
    \centering
\includegraphics[width=\textwidth]{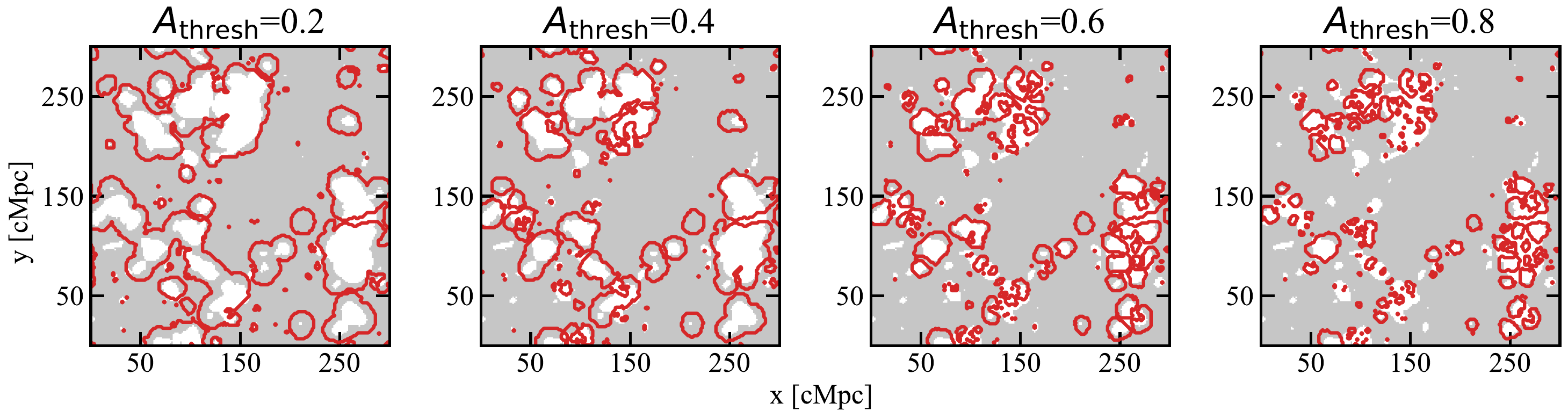}
    \caption{Recovered bubble maps using different $A_{\rm thresh}$. From left to right column are the maps using $A_{\rm thresh}$ of 0.2,0.4,0.6,0.8, respectively. When $A_{\rm thresh}>0.5$, the recovered bubbles break into too many small pieces.}
    \label{fig:asym_map_vali_examp}
\end{figure*}

Figure~\ref{fig:asym_map_vali_examp} shows the asymmetry maps made using different $A_{\rm thresh}$ for a window radius of 12 cMpc and depth of 4 cMpc (see Section~\ref{sec:asymscoremethod} for a description of the window). As $A_{\rm thresh}$ increases, the recovered bubbles become more fractured. This is because the asymmetry scores around bubble edges are not constant. Only a few areas around bubble edges have very high asymmetry scores. If we only connect these high asymmetry score points, the recovered bubble becomes small and fractured. We find $A_{\rm thresh}>$0.5 will result in too many small bubbles. 

We test for the best $A_{\rm thresh}$ using our $z=8$ simulations, assuming we can detect all the galaxies with $\MUV\leq-17.8$ (corresponding to 0.002 galaxies/cMpc$^3$) and a 5$\sigma$ \lya EW detection limit of 30\,\AA\ (an optimistic observational requirement, see Appendix~\ref{app:obs_asym_ngal}). We vary $A_{\rm thresh}$ between 0.15 and 0.4 and use a combination of visual and the Jaccard index (see Appendix ~\ref{sec:bestmetric}) to find the best $A_{\rm thresh} \approx 0.3$. 
We find this $A_{\rm thresh}\approx0.3$ is independent of the IGM neutral fraction for $0.4\leq\xHI\leq0.9$, and does not depend strongly on the window size (Appendix~\ref{sec:bestwindowrd}), making it a robust, model-independent, choice for making bubble maps.

\subsection{Comparison metrics}\label{sec:bestmetric}

To evaluate the recovery of the bubble maps relative to the input ionization map we use the Jaccard index \citep[$\mathcal{J}$,][]{Jaccard1912}. This is defined as the fraction of correctly recovered ionized/neutral pixels in the map:
\begin{equation}
\mathcal{J}\equiv\frac{\rm{number\,of\,correctly\,recovered\,neutral\,or\,ionized\,pixels}}{\rm{number\,of\,pixels\,in\,the\,asymmetry\,map}}
\end{equation}

The Jaccard index directly reflects how good our recovered map is compared to the input one. $\mathcal{J}=0$ ($\mathcal{J}=1$) means the recovered map is completely different from (the same as) the input map. By visual inspection of a range of maps at different \xHI from our simulations, we found $\mathcal{J}>0.8$ indicates a good recovery of bubble maps.

\begin{figure*}
    \centering
    \includegraphics[width=\textwidth]{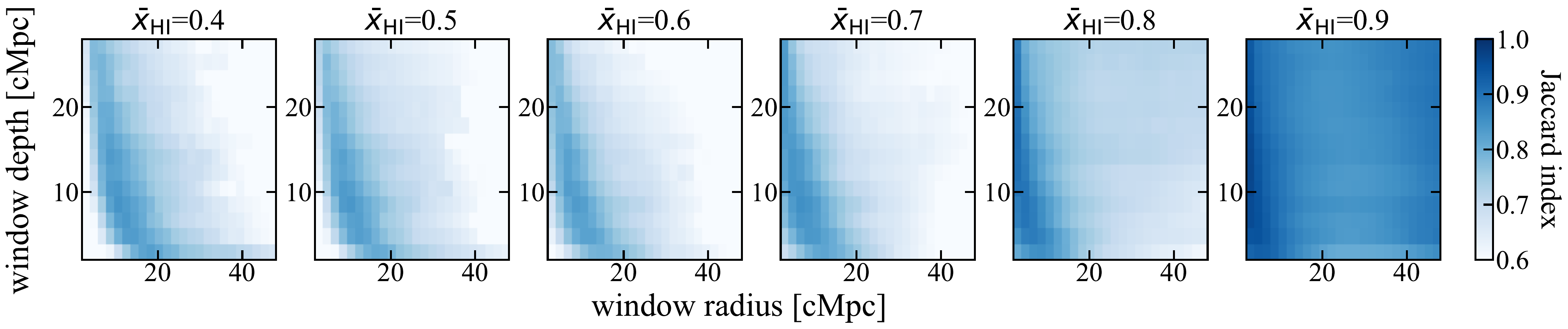}
    \caption{Jaccard index for different window radius and depth combinations as a function of \xHI.}
    \label{fig:3like_grid}
\end{figure*}

In Figure~\ref{fig:3like_grid} we show the values of Jaccard index for different window radius and depth combinations at \xHI=0.4-0.9. The Jaccard index clearly traces the optimum parameters until $\xHI=0.9$, when the mean bubble size becomes small and most of the field is neutral. Thus we use a combination of the Jaccard index with visual inspections of the maps to establish the best parameters for our method.

\subsection{Optimal window shape}\label{sec:bestwindowrd}
To determine the best window radius and depth combination for the asymmetry score (Equation~\ref{eq:asymvec}), we calculate the Jaccard index for test window radii and depths for simulations at \xHI=0.4-0.9 (Figure~\ref{fig:3like_grid}). We fix the $A_{\rm thresh}=0.4$. We find the best recovery is for window radius $<$30 cMpc and window depth$<$25cMpc. For window depth $<$30cMpc the best window radius and depth are anti-correlated. This is because as the window depth increases, the true bubbles in the stacked ionization map become smaller, a small window radius is optimal to resolve small true bubbles. The window depth needs to be thin enough to prevent ionized bubbles from being washed out in the stacked ionization maps, but thick enough to include sufficient to sample the space.
By visual inspection, we find the best parameters for our entire test \xHI range are [window radius, window depth]=[12, 4]cMpc.

\section{Tests of observational requirements for asymmetry maps} \label{app:obs_asym}

Here we describe the number density of sources and \lya EW limit required to map bubbles in the plane of the sky.

\subsection{Required galaxy number density}  \label{app:obs_asym_ngal}
To test the required number density of sources for creating robust asymmetry maps, we create a series of mock surveys at $z=8$ varying the UV magnitude limit, $M_{\rm UV, limit}$, of detectable sources, assuming a 5$\sigma$ \lya EW detection limit of 30\,\AA\ for the faintest sources. We infer \lya transmissions of individual galaxies using the Bayesian method as described in Section~\ref{sec:galmod}, using the maximum likelihood value of $\mathcal{T}$. We use a combination of visual inspection and the Jaccard index to judge how well the input bubble maps are recovered. 

\begin{figure*}[h]
    \centering
    \includegraphics[width=0.85\textwidth]{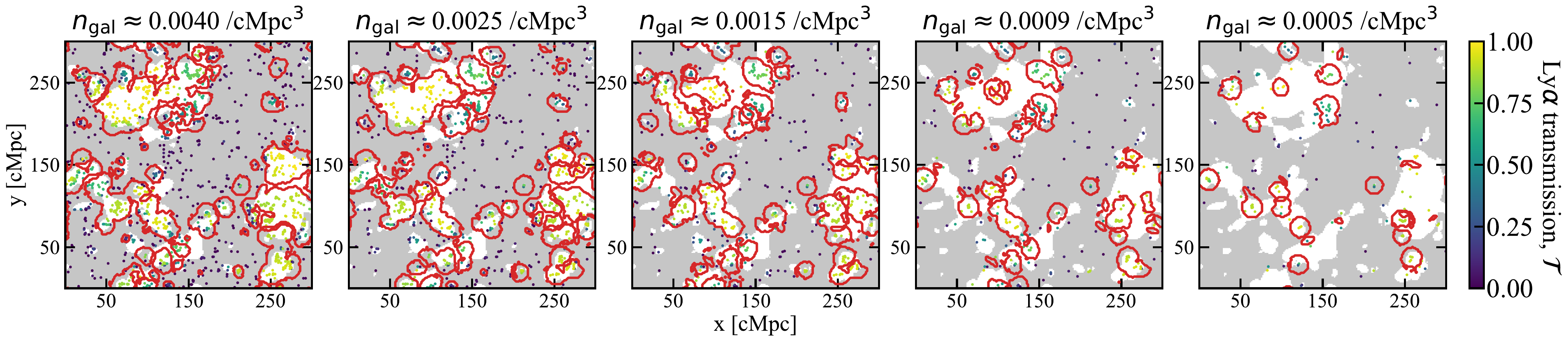}
    \caption{Recovered bubble maps corresponding to different galaxy number densities $n_{\rm gal}\approx0.0040-0.0005$/cMpc$^{3}$ at \xHI=0.7. To achieve these number densities we vary $M_{\rm UV, limit}$ in our $z=8$ simulations, and assume flux limit corresponding to a 5$\sigma$ EW limit of 30\,\AA\, for the faintest galaxies. Dots show the detectable galaxies, color-coded by their true \lya transmissions. As number density of galaxies increases, bubble recovery improves. With $n_{\rm gal}\simgt$0.002 /cMpc$^{3}$ we can recover bubbles well: lower $n_{\rm gal}$ is not sufficient to sample the space.}
    \label{fig:Mlim_recovered}
\end{figure*}

Figure~\ref{fig:Mlim_recovered} shows the recovered bubble maps for $n_{\rm gal}\approx$ 0.0040, 0.0025, 0.0015, 0.0009, and 0.0005 /cMpc$^3$ at \xHI=0.7. We find excellent recovery of bubbles for $n_{\rm gal} \geq 0.002$/cMpc$^3$. Lower galaxy number densities do not sufficiently sample the space, resulting in recovered bubbles only centered around the strongest galaxy overdensities.

\subsection{Required EW limit} \label{app:obs_EW}

\begin{figure*}[h]
    \centering
    \includegraphics[width=0.7\textwidth]{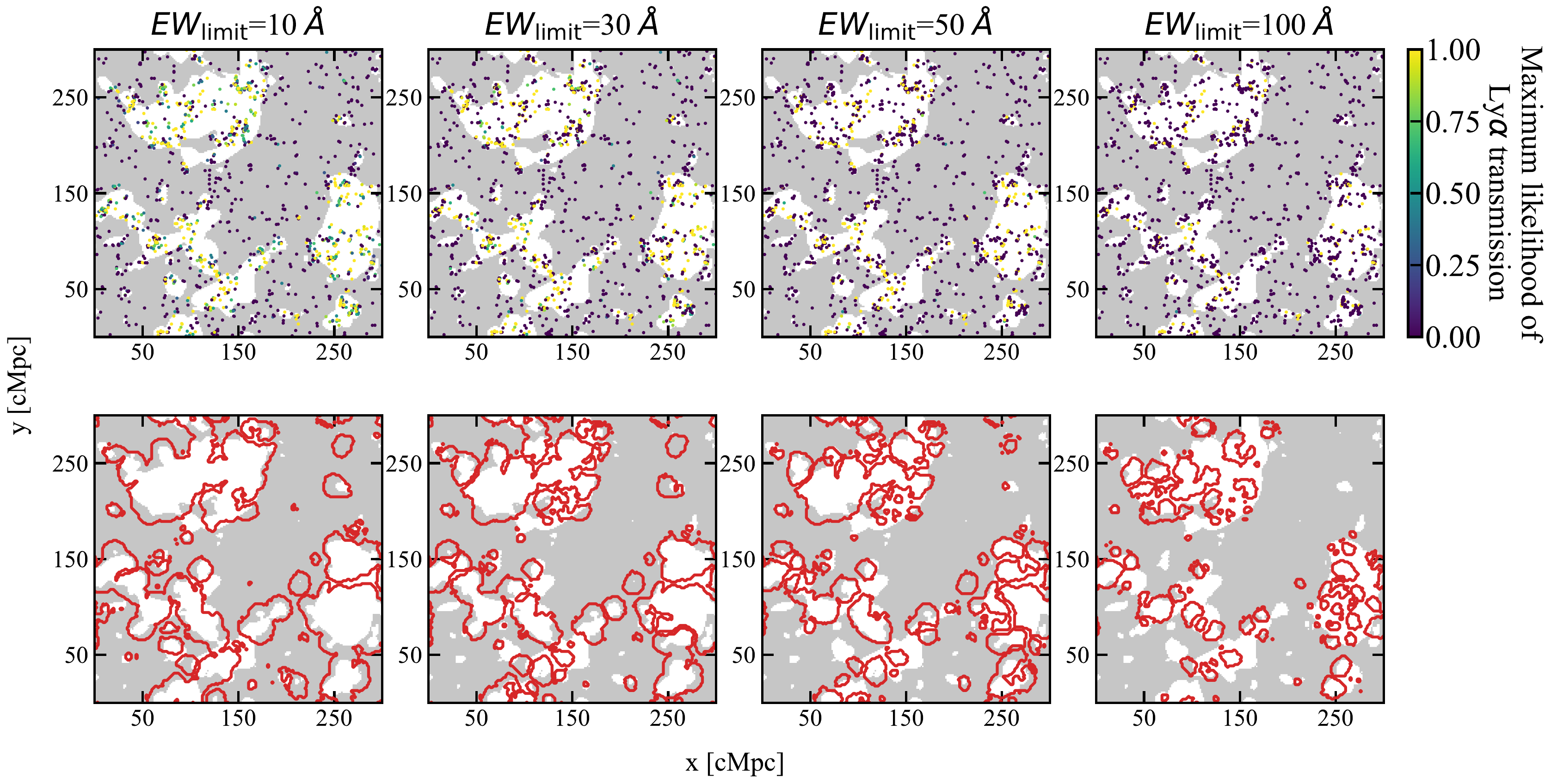}
    \caption{\textit{Upper panel:} Input ionization map (grey neutral, white ionized) at \xHI=0.7 and \MUV$\leq$-17.2 galaxies (dots) used to calculate asymmetry maps at $z=8$. Galaxies are color-coded by their peak \lya transmission ($\mathcal{T}$) inferred from the observed EW and uncertainty of EW. From left to right we show results using 5$\sigma$ EW limit, $EW_{\rm Ly\alpha, limit}$= 10, 30, 50, and 100 \AA\ for the faintest galaxies. \textit{Lower panel:} 
    Asymmetry maps (red contours) made using the galaxies in the upper panel. We can recover bubble edges very well when $EW_{\rm Ly\alpha, limit}$ is at least the median emergent EW of galaxies ($\approx 30$\,\AA). When $EW_{\rm Ly\alpha, limit}$ is too shallow ($\gg$30\,\AA), we cannot get useful information about \lya transmission from our observations, thus the recovered bubbles only trace the regions around galaxies with EW greater than the detection limit.}
    \label{fig:Flim_asymmap_05}
\end{figure*}

To estimate the EW limit requirements for mapping bubbles using the asymmetry method, we create mock surveys at a range of \xHI now varying the \lya EW limit, $EW_{\rm Ly\alpha, limit}$ for the faintest observable galaxies (i.e. we vary the flux limit, so brighter galaxies will have lower EW limits), assuming $n_\mathrm{gal}=0.004/$cMpc$^3$ ($M_{\rm UV, limit}=-17.2$ at $z=8$). Figure~\ref{fig:Flim_asymmap_05} shows example asymmetry maps at \xHI=0.7 using $EW_{\rm Ly\alpha, limit}=10-100$\,\AA\ for the faintest galaxies.

As expected, we can recover bubble edges very well for \lya EW limits deeper than the median pre-IGM damping \lya EW ($\approx30$\AA, see Section~\ref{methods:T}). The shallower the \lya EW limit is, the more likely we break big ionized bubbles into small bubbles around high EW sources, as we lose information about $\mathcal{T}$. 

\begin{figure}
    \centering
    \includegraphics[width=0.7\columnwidth]{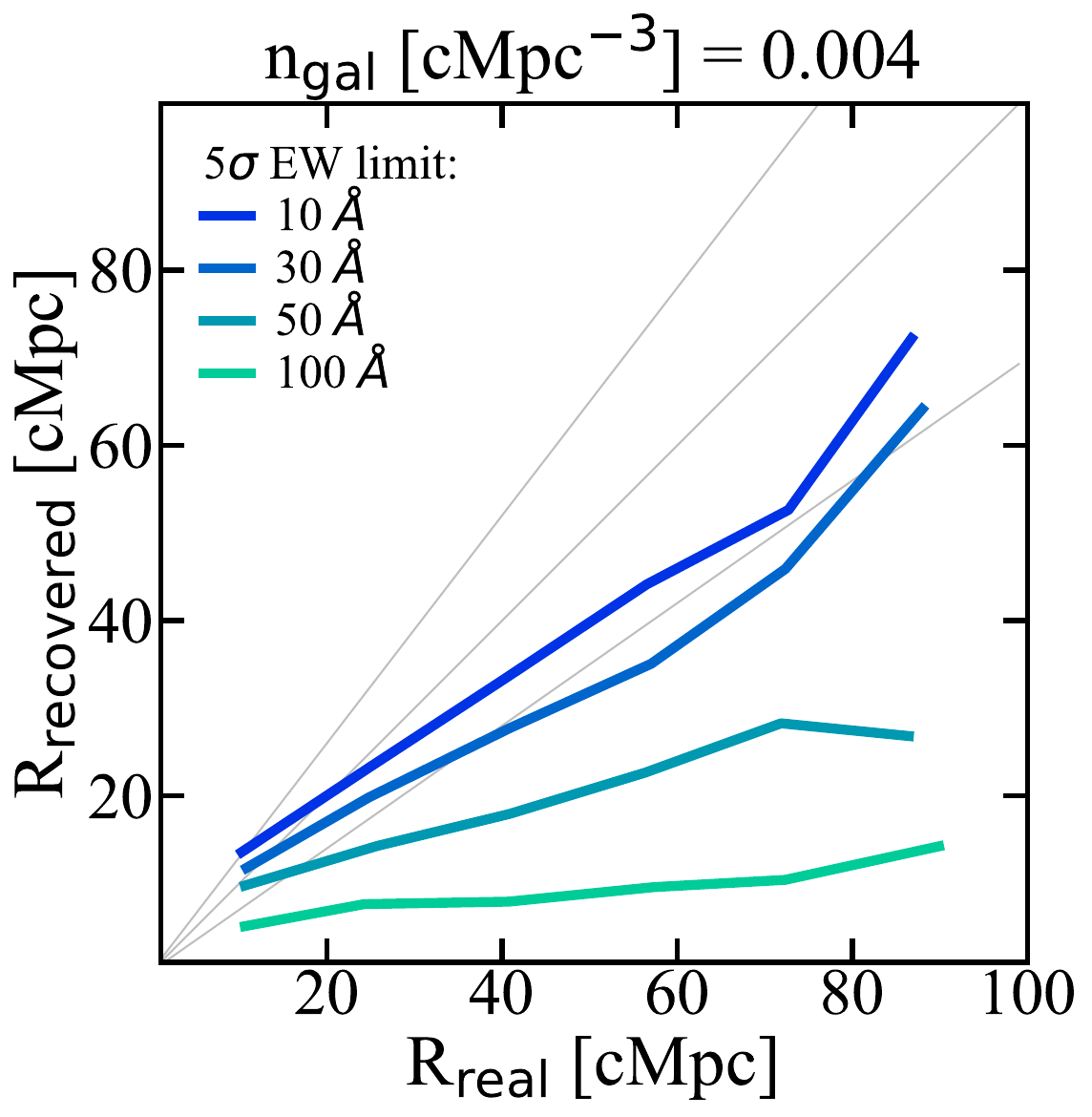}
    \caption{Median recovered bubble size versus real bubble size for 5$\sigma$ \lya equivalent width limit = [10, 30, 50, 100] \AA, galaxy number density $\approx$0.004/cMpc$^{3}$ in our asymmetry method. Grey lines from top to bottom show $+30\%$, 0$\%$, and $+30\%$ recover bubble size error, respectively. We can recover bubble size to within 30$\%$ error when galaxy number density is greater than 0.004/cMpc$^{3}$.}
    \label{fig:Rrec_asymmap}
\end{figure}

We test the bubble size recovery by randomly selecting positions where both the recovered bubble map and the input bubble map are ionized, measuring the sizes of bubbles around the points using the mean free path method \citep{Mesinger2007}, and comparing the recovered bubble sizes to the simulated bubble sizes. We find our asymmetry method can recover bubble sizes to within $30\%$ of the true size when $n_{\rm gal} \geq$0.004 for 5$\sigma$ \lya EW detection limit $\lesssim$ 30\,\AA\, for 10 cMpc $\lesssim R\lesssim$ 100 cMpc bubbles (Figure~\ref{fig:Rrec_asymmap}). We slightly overestimate R$\lesssim$10 cMpc bubbles because the spatial resolution is restricted by our window radius. We slightly underestimate $R\gtrsim$60\,cMpc bubbles as very large bubbles can be split by our algorithm - further refinement of $A_\mathrm{thresh}$ could improve the segmentation of bubbles. 

\end{appendix}
\end{document}